\documentclass[10pt,conference]{IEEEtran}

\AtBeginDocument{%
  \providecommand\BibTeX{{%
    \normalfont B\kern-0.5em{\scshape i\kern-0.25em b}\kern-0.8em\TeX}}}

\usepackage{soul}
\usepackage{float}
\usepackage{threeparttable}
\usepackage{booktabs}
\usepackage{multirow}
\usepackage{epstopdf}
\usepackage{hyperref}
\usepackage{listings}
\usepackage{fancybox}
\usepackage{graphicx}
\usepackage{subcaption}
\usepackage{paralist}
\usepackage{ragged2e}
\usepackage{enumitem}
\usepackage[table]{xcolor}
\usepackage{xcolor}
\usepackage[normalem]{ulem} % use normalem to protect \emph
\newcommand\hlg{\bgroup\markoverwith
  {\textcolor{green!10}{\rule[-.5ex]{2pt}{2.5ex}}}\ULon}
\newcommand\hlb{\bgroup\markoverwith
  {\textcolor{blue!10}{\rule[-.5ex]{2pt}{2.5ex}}}\ULon}
\newcommand\hlr{\bgroup\markoverwith
  {\textcolor{red!10}{\rule[-.5ex]{2pt}{2.5ex}}}\ULon}

\usepackage[framemethod=TikZ]{mdframed}
\usepackage{tcolorbox}% http://ctan.org/pkg/tcolorbox
\makeatletter
\newcommand{\mybox}[1]{%
  \setbox0=\hbox{#1}%
  \setlength{\@tempdima}{\dimexpr\wd0+13pt}%
  \begin{tcolorbox}[boxrule=0.5pt, colback=white, arc=4pt,
      left=6pt,right=6pt,top=6pt,bottom=6pt,boxsep=0pt]
    #1
  \end{tcolorbox}
}
\usepackage{tikz}
\usepackage{pgfplots}
\usetikzlibrary{pgfplots.statistics,calc}
\usepackage{color}
\usepackage{xcolor}
\usepackage{array}
\usepackage{amsmath}
\usepackage{centernot}
\usepackage{xspace}
\usepackage{url}
\usepackage{verbatim}
\usepackage{wrapfig}
\usepackage{tabularx}
\usepackage{bbding}
\usepackage{pifont}
\usepackage{wasysym}
\usepackage{amssymb}

\newcommand{\tool}{\texttt{QTypist}}
\makeatletter  
\newif\if@restonecol  
\makeatother  
\usepackage[linesnumbered,ruled,vlined]{algorithm2e}%[ruled,vlined]{  
\usepackage{algpseudocode}  
\usepackage{amsmath}  
\begin{document}


%\title{QTypist: Intelligent Textual Input Generation for Mobile GUI Testing}
% \title{Fill in the Blank: Automated Textual Input Generation for Mobile GUI Testing}
\title{Fill in the Blank: Context-aware Automated Text Input Generation for Mobile GUI Testing}

\author{\IEEEauthorblockN{1\textsuperscript{st} Zhe Liu}
\IEEEauthorblockA{\textit{Institute of Software} \\ \textit{Chinese Academy of Sciences}, \\ Beijing, China \\
liuzhe181@mails.ucas.edu.cn
}
\\
\IEEEauthorblockN{4\textsuperscript{th} Xing Che, \\ 5\textsuperscript{th} Yuekai Huang}
\IEEEauthorblockA{\textit{Institute of Software} \\ \textit{Chinese Academy of Sciences}, \\ Beijing, China}
\and
\IEEEauthorblockN{2\textsuperscript{nd} Chunyang Chen}
\IEEEauthorblockA{\textit{Monash University}, \\ 
Melbourne, Australia\\
Chunyang.chen@monash.edu
\\ \ 
}
\\
\IEEEauthorblockN{6\textsuperscript{th} Jun Hu}
\IEEEauthorblockA{\textit{Institute of Software} \\ C\textit{hinese Academy of Sciences}, \\ Beijing, China \\
hujun@iscas.ac.cn}
\and
\IEEEauthorblockN{3\textsuperscript{rd} Junjie Wang}
\IEEEauthorblockA{\textit{Institute of Software} \\ \textit{Chinese Academy of Sciences} \\
\textit{*Corresponding author} \\
junjie@iscas.ac.cn}
\\
\IEEEauthorblockN{7\textsuperscript{th} Qing Wang}
\IEEEauthorblockA{\textit{Institute of Software} \\ \textit{Chinese Academy of Sciences} \\
\textit{*Corresponding author} \\
wq@iscas.ac.cn}
}

\maketitle

\begin{abstract}
% Automated GUI testing is an important way to ensure the quality of mobile app. However, most of them focus on exploration algorithm improvement, while few of them are concerned with the complicated interaction with app like text input generation.

Automated GUI testing is widely used to help ensure the quality of mobile apps. 
However, many GUIs require appropriate text inputs to proceed to the next page, which remains a prominent obstacle for testing coverage.
Considering the diversity and semantic requirement of valid inputs (e.g., flight departure, movie name), it is challenging to automate the text input generation.
Inspired by the fact that the pre-trained Large Language Model (LLM) has made outstanding progress in text generation, we propose an approach named {\tool} based on LLM for intelligently generating semantic input text according to the GUI context.
To boost the performance of LLM in the mobile testing scenario, we develop a prompt-based data construction and tuning method which automatically extracts the prompts and answers for model tuning.
We evaluate {\tool} on 106 apps from Google Play, and the result shows that the passing rate of {\tool} is 87\%, which is 93\% higher than the best baseline. We also integrate {\tool} with the automated GUI testing tools and it can cover 42\% more app activities and 52\% more pages compared with the raw tool. 
\end{abstract}


\section{Introduction}
\label{sec_introduction}
% \chen{Do we need to mention GUI? Actually it is just mobile testing, right?}
Due to the portability and convenience of mobile phones~\cite{Number2020Google}, mobile applications (apps) now have become indispensable for our daily life in accessing the world~\cite{Mobile2016first,jabbarvand2019search,matinnejad2017automated}. 
% The importance of mobile apps also makes it vital for the development team to carry out a through testing for ensuring the quality of mobile apps, which decides its success among many similar apps in the market. 
However, it is challenging to guarantee the app quality, especially considering that mobile apps interact with complex environments (e.g., users, devices, and other apps). 
Since GUI provides a bridge between software apps and end-users through which they can interact with each other, GUI testing is widely used to test if the application is functioning correctly. 
% User Interfaces, as the primary interface of user-app interactions, are natural entry points for app testing.
Although manual GUI testing is often used in practice, automated GUI testing is becoming popular to save human efforts and scaled to different apps on different devices. 
% \chen{Since GUI testing is well established, you can simplify the first paragraph.}
% \chen{Need to explicitly emphasize the importance of testing coverage}

\begin{figure}[htb]
\centering
\vspace{-0.05in}
\includegraphics[width=8.6cm]{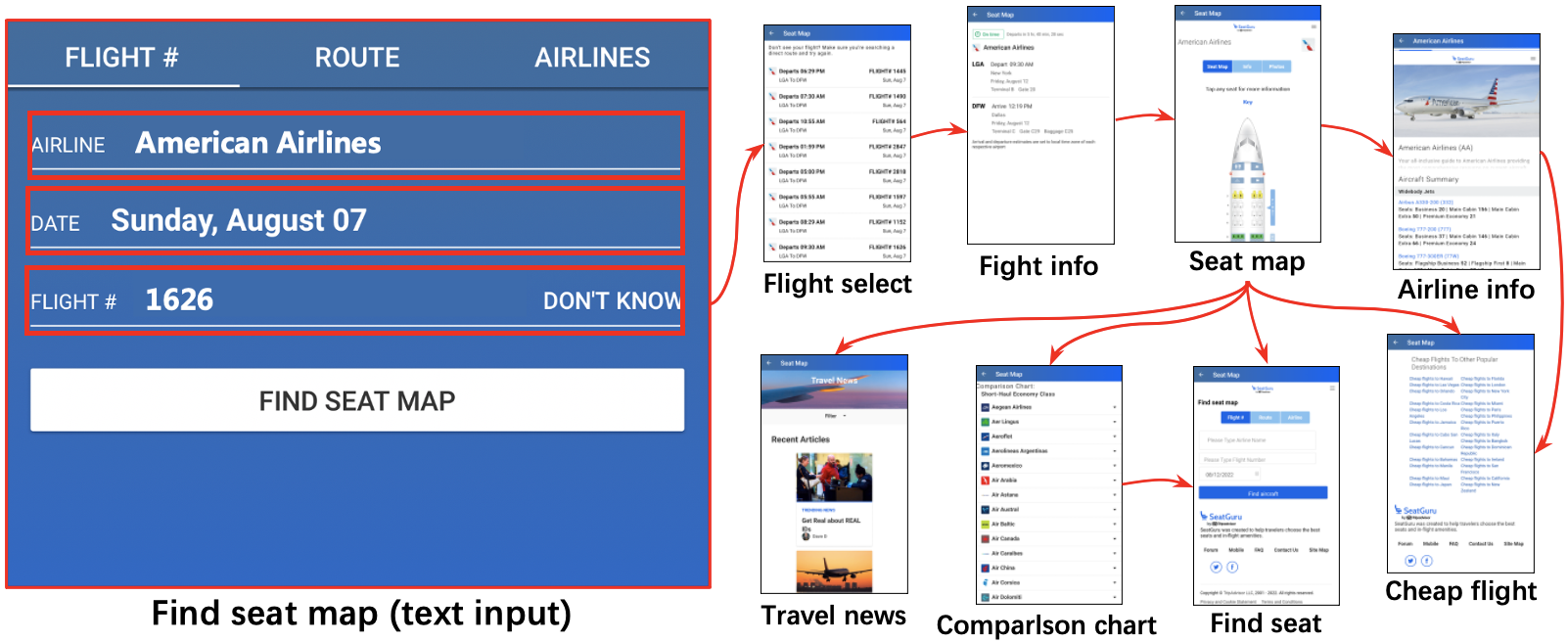}
% \vspace{-0.1in}
\caption{Example of text input in Android app. 
% The red arrow represents UI page that requires correct text input to cover.
}
\label{fig:Example-input-ATG}
\vspace{-0.1in}
\end{figure}

There are many automated mobile GUI testing approaches, including model based~\cite{mirzaei2016reducing,yang2018static,yang2013grey}, probability based~\cite{machiry2013dynodroid,zeng2016automated,mao2016sapienz} and deep learning based~\cite{li2019humanoid,pan2020reinforcement} ones to dynamically explore mobile apps by executing different actions such as scrolling, clicking based on the analysis of code structure of the current page to verify UI functionality.
However, most of them focus on exploration algorithm improvement, while few of them are concerned with the complicated interaction with app like \textit{text input} generation.
According to our observation in Section~\ref{sec_motivation}, most apps have some pages requiring specific text inputs to go to the next page.
As seen in Fig~\ref{fig:Example-input-ATG}, without correct information for flight searching, the consecutive pages such as search display, flight info, travel news, airline info and seat map cannot be accessed.
That low activity coverage will negatively affect the testing adequacy of automated GUI testing tools~\cite{wang2018empirical,wang2021vet,choudhary2015automated,su2021benchmarking,fan2018large}. 
% That low activity coverage will negatively affect the ability of automated GUI testing tools to detect bugs~\cite{wang2018empirical,wang2021vet,choudhary2015automated,su2021benchmarking,fan2018large}. 

\begin{figure*}[t]
\centering
\vspace{-0.12in}
\includegraphics[width=18cm]{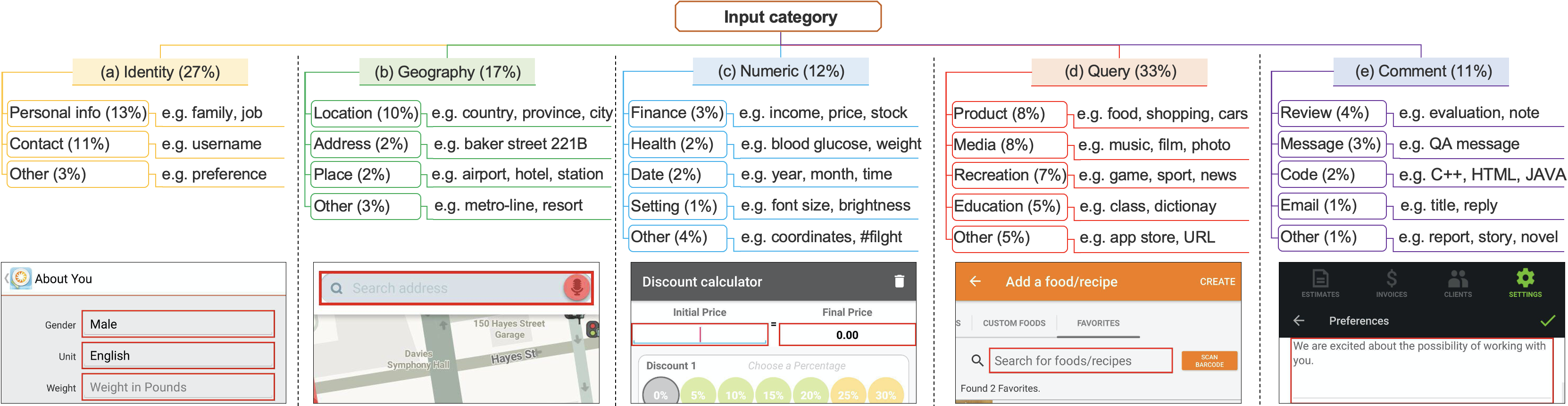}
% \vspace{-0.05in}
\caption{Example of 5 categories of text input.} 
\label{fig:5-kind-input}
\vspace{-0.1in}
\end{figure*}

Text input generation is a challenging task to automated GUI testing, even to humans~\cite{su2021benchmarking,he2020textexerciser,liu2017automatic,wang2021infrastructure,kong2018automated,rubinov2018we,pecorelli2022software}.
First, it requires a specific value for different types of the inputs, such as weight for identify setting, the street address for map apps, blood pressure for health apps, movie name for media apps, and source code for coding app (more details seen at Section~\ref{sec_motivation_types_UIs}).
Second, some text inputs within the same UI page may correlate with each other such as different locations in departure and arrival in flight searching, and minimum price setting smaller than that of maximum price in the filter for product searching.
Without valid inputs, automated testing tools are unable to proceed to the next UI page, resulting in low testing adequacy and missing potential bugs.

Most current approaches either generate the text inputs with fixed/random string or words like ``hello world''~\cite{zeng2016automated,machiry2013dynodroid,azim2013targeted,gu2019practical,li2017droidbot}, or limited heuristic rules~\cite{he2020textexerciser,trinh2014s3,anand2012automated,arnatovich2018mobolic,trinh2014s3,sunman2022automated} by analyzing the source code and extracting constraints.
Towards generating semantic input, Liu et al.~\cite{liu2017automatic} proposed an RNN (Recurrent Neural Network) based method incorporated with Word2Vec.
But it needs a lot of manually labeled data, which can not be generalized to different apps, and does not consider the contextual information mentioned above.

The emerging Large Language Model (LLM)~\cite{attention,brown2020GPT3,DBLP:journals/jmlr/RaffelSRLNMZLL20,devlin2018bert} trained on ultra-large-scale corpus is potentially able to understand the input prompt and generate reasonable output.
For example, GPT-3~\cite{brown2020GPT3} (Generative Pre-trained Transformer-3) from OpenAI with 175 billion parameters trained on billions of tokens can smartly generate articles~\cite{lucy2021gender,sharples2022automated}, answer questions~\cite{yang2022empirical}, write source code~\cite{macneil2022generating}.
Inspired by the power of LLM, we formulate text input generation question into a close-style fill-in-blank questions answering task by regarding LLM as a knowledge base. 
Once UI page requiring text inputs is passed, more automated testing techniques can be applied to detect bugs~\cite{su2017guided,li2017droidbot,cai2020fastbot,gu2019practical,pan2020reinforcement}.

Inspired by the fact that the LLM has made outstanding progress in news generation, email reply, abstract extraction, etc.~\cite{brown2020GPT3,laskin2020reinforcement,yang2022empirical,chen2020big},
% input widgets often have a context relationship with the information on GUI pages and it needs to have practical meaning,
we propose an approach, {\tool}\footnote{Our approach is named as {\tool} as it is like an intelligent typist writing correct input text according to different mobile app input scenarios.} to model the text input information by pre-trained Large Language Model (LLM) to automatically generate input text. Given a GUI page with text input and its corresponding view hierarchy file, we first extract the context information for the text input, and design linguistic patterns to generate prompts for inputting into the LLM. To boost the performance of LLM in mobile input scenarios, we develop a prompt-based
data construction and tuning method, which automatically builds the prompts and answers for model tuning.
% \rev{Note that one strength of our approach over conventional program analysis is that it does not need app source code and can be applied to many languages, including English, Korean, French, etc.}

To evaluate the effectiveness of {\tool}, we carry out an experiment on 168 text inputs (i.e., EditText) from 106 popular Android apps in Google Play.
Compared with 11 common-used and state-of-the-art baselines, {\tool} can achieve more than 93\% boost in passing rate compared with the best baseline, resulting in 87\% passing rate.
As {\tool} can also generate text with actual meaning, we carry out a user study to check its diversity and significance, the results demonstrate that the text generation performance average score is 4.3 (out of 5). 
Apart from the accuracy of {\tool}, we also evaluate the usefulness of {\tool} by integrating it with 3 commonly-used and state-of-the-art automated GUI testing tools. The result shows that the Ape, DroidBot and Monkey with integrated {\tool} covers 42\%, 33\% and 28\% more activities and 52\%, 41\% and 30\% more UI pages. 
% Compared with using Ape alone, the Ape with {\tool} detects 122\% more bugs, thanks to the more covered UI pages by the automated input generation.
We hope that our results can raise the community's awareness of complex operations like text input beyond the community's existing heavy focus on exploration algorithms.

The contributions of this paper are as follows:
\begin{itemize}
\item The first work to formulate the text input generation problem as a cloze-style fill-in-blank language task to assist existing GUI testing tools in achieving higher testing coverage.
%which motivates follow-up studies in this task, and use the popular ``pre-train, prompt and predict'' paradigm to solve the task.

% 第一个关于输入组件的经验研究
\item The empirical investigation and categorization of text inputs, which helps to understand the text input generation task, as well as a public available dataset of diversified mobile app text inputs for follow-up studies.
% \rev{, which has five categories text inputs that can help understand the text input generation.}

% 提出了新的模式（参考邢老师的文章）
% \item This is the first work to formulate the input content inference problem as a cloze-style fill-in-blank language task, and propose a novel ``pre-train, prompt and predict'' paradigm to solve the task.

% 我们的生成模型
\item A novel approach {\tool} based on ``pre-train, prompt and predict'' paradigm of the large language model by understanding the local and global context for automatically inferring semantic text input. 
%provides the dataset, source code of {\tool}, and the detailed experimental results of this paper.\label{github}} 
% for input text generation, 
%which models text input information by pre-trained Large Language Model to automatically generate input text. It contains a context-aware prompt generation method which extracts and organizes information of text inputs, to stimulate pre-trained model to understand text inputs and generate correct output. 
% \rev{We propose a context aware prompt generation method to stimulate the pre-trained model to understand text inputs' contextual information. We design a prompt-based data construction and tuning method for model prompt tuning. }

% % 评估情况
% \item Effectiveness evaluation of {\tool} with high state transition pass rate compares with baselines. Our approach achieves superior performance even with only a small amount of prompt learning data. The case study demonstrates its potential usefulness in augmentation with automated GUI testing.
\end{itemize}

% The contributions of this paper are as follows:
% \begin{itemize}
% % 第一个关于输入组件的经验研究
% \item This is the first work to conduct a systematical investigation of input text in real-world mobile apps. We a dataset of different app input text and release it for follow-up studies.

% % 提出了新的模式（参考邢老师的文章）
% \item This is the first work to formulate the input content inference problem as a cloze-style fill-in-blank language task, and propose a novel ``pre-train, prompt and predict'' paradigm to solve the task.

% % 我们的生成模型
% \item A novel approach {\tool}\footnote{\url{https://github.com/testgod6/typist} provides the dataset, source code of {\tool}, and the detailed experimental results of this paper.\label{github}} for input text generation, which models the input widget information by Generative Pre-Train network to automatically predict the input text. 

% % 我们的提示学习方法
% \item We also propose the first prompt learning method to stimulate the pre-trained model to understand the input widget' semantic and contextual information. We design a context aware prompt generation method and a heuristic-based answer construction method for prompt tuning. 

% % 评估情况
% \item Effectiveness evaluation of {\tool} with high state transition pass rate compares with baselines. Our approach achieves superior performance even with only a small amount of prompt learning data. The case study demonstrates its potential usefulness in augmentation with automated GUI testing.
% \end{itemize}

\section{motivational study and background}
\label{sec_motivation}
% \chen{I do not know why there is no empty space at the beginning of each paragraph. Please follow the template, otherwise may be desk rejected.}
To better understand the text inputs in real-world practice, we carry out a pilot study to examine their prevalence. 
The pilot study also explores what categories of input text exist, so as to facilitate the design of our approach for extracting input information and generating text.

\subsection{Motivational Study}
\label{sec_motivation_data_collection}
\subsubsection{\textbf{Data Collection}}
Our experimental dataset is collected from one of the largest Android UI datasets Rico~\cite{Rico}, which has a great number of Android UI screenshots and their corresponding view hierarchy files.
% \footnote{http://interactionmining.org/rico\#quick-downloads}.
These apps belong to diversified categories such as news, entertainment, medical, etc. 
% The reason why we utilize this dataset is that it includes both the UI screenshots and the corresponding view hierarchy files which facilitates the searching and analysis of input text.
We analyze the view hierarchy file according to the package name and extract GUI page belonging to the same app. 
A total of 7,000 apps with each having more than 3 pages are extracted through the above methods. % are selected in this study.
For the selected apps, we first randomly select 50 apps with 216 GUI pages and check their text inputs through view hierarchy files. 
We summarize a set of keywords that the apps have text inputs~\cite{Textinput} (UI elements for entering text)
% (UI elements for entering and modifying text), 
% \jie{what is modifying text?}
% \chen{Great to give a formal definition.}
e.g., \textit{EditText, hint-text, AutoCompleteTextView, etc}\textsuperscript{\ref{github}}.
We then use these keywords to automatically filter the view hierarchy files from the remaining 6,950 apps.
After that, we obtain 5,663 (80.9\%) candidate apps with potential text inputs. Four authors then manually check them to ensure that they have text inputs. The fifth author reexamines the results until a consensus is reached.
Finally, a total of 5,066 (73.8\%) apps are identified to have text inputs.
% \jie{comment}
%这里其实不是说这个比例的app含有input，因为你前面把少于三个page的去掉了是吧？所以我后面用了about
% \liuzhe{comment}
% 嗯嗯是的

% \subsection{How does Textual Input Affect Automated GUI Testing?}
% \label{sec_motivation_types_number}
% \liuzhe{
% To understand the proportion of textual inputs in the app and the impact on activity coverage, four authors then manually check the above-returned candidate apps to ensure that they are truly related to the input widget. The fifth author reexamines the results until a consensus is reached.
% Finally, a total of 5,066 (73.8\%) apps were identified to have textual inputs. \chen{How to calculate this?}
% This result indicates that the textual inputs account for a non-negligible portion of mobile apps and should be paid careful attention for improving the software quality.
% We further analyze the activities related to the input text and randomly select 150 apps (can install) for each category. Four authors manually label the activity transition graph (ATG) \chen{Not 100\% accurate} of them, and label the activities that need to be explored after entering the correct text. The experimental results show that the average number of activities that can only be explored after the correct textual input accounts for an average 63.6\% of the total number of activities.}
% \chen{I feel that we do not need this subsection very much.}
% \liuzhe{comment}
% 这里想说一下有多少存在输入组件的app，（关键词筛选完人工又标注了一边）
% app输入组件之后大概有%的activities
% 感觉这两部分可以合并在上面一段和下面一段

% \input{figure/input-cate}

% \subsection{Categorizing Input Text}
\subsubsection{\textbf{The Categories of Text Inputs}}
\label{sec_motivation_types_UIs}
% \rev{Since the target of this work is to generate the input text, we further analyze the apps with text input in Section} \ref{sec_motivation_data_collection}. 
% \jie{comment}
%这里最后一句不是已经说了，得到的5066个apps都是有text input的了。我觉得这句话就可以删掉了。
We randomly select 2,600 apps with pages requiring text input and conduct manual categorization to derive the type of input widget according to the input content. 
%Two authors, who have more than four years of Android development experience, jointly participate in the process. 
%Following the open coding protocol~\cite{seaman1999qualitative}, we analyze the apps and categorize the input texts.
Following the open coding protocol~\cite{seaman1999qualitative}, two authors with more than four years of Android development experience individually examine the content of the text input, including the app name, activity name, input type and input content.
Then, each annotator iteratively merges similar codes, and any disagreement of the categorization will be handed over to the third experienced researcher for double checking. 
%We group similar codes into one category, and the grouping process is iterative. 
%Specifically, we constantly move back and forth between the category/sub-category and input widget to improve the classification. 
%In the absence of an agreement between the two authors, the other authors act as arbitrators to discuss and resolve the conflict.
%Specifically, for those textual inputs assigned with different labels, the four authors re-identify the categories and determine whether new categories/sub-categories are needed.
%We follow the procedure until all authors reach an agreement.
Finally, we construct a taxonomy of text inputs in Fig~\ref{fig:5-kind-input} including 5 main categories 
% including \textit{personal information, geographic information, numerical information, search content} and \textit{comment content} with details 
as follows. 
% \jie{Note that, some input widgets can involve }

% \textbf{Personal information (27\%)}: 
\textbf{Identity (27\%)}: 
As shown in Fig \ref{fig:5-kind-input} (a), the input text is usually related to personal information, such as user name, family, email, job, etc. It usually appears with \textit{MainAcitivity}, \textit{ContactActivity}, \textit{PersonalActivity}, etc.

% \textbf{Geographic information (17\%)}: 
\textbf{Geography (17\%)}: 
As shown in Fig \ref{fig:5-kind-input} (b), the input text is usually related to the navigation \& travel apps, such as location, address and country, etc.
This category may share contextual constraints, such as different addresses of departure and arrival.

% \textbf{Numerical information (12\%)}: 
\textbf{Numeric (12\%)}: 
As shown in Fig \ref{fig:5-kind-input} (c), the input text is usually numbers, such as year, weight, income, blood glucose, etc.
% \rev{It is necessary to consider the constraints between different text inputs.}
These text fields are often correlated, e.g., the value of a ``maximum wage'' field should be larger than the one of a ``minimum wage''.

\textbf{Query (33\%)}: 
As shown in Fig \ref{fig:5-kind-input} (d), the input text is often meaningful, such as game, shopping, music, film, etc.
It is necessary to consider the semantic information of the text inputs. e.g., for a game download app, you need to enter a game name.

% \textbf{Comment content (11\%)}:
\textbf{Comment (11\%)}:
As shown in Fig \ref{fig:5-kind-input} (e), the input text are mainly
% \rev{is often large, such as} 
reviews, remarks, notes, diary, code, etc.
It usually needs to consider the category of the app, e.g., the code compiler needs to enter code information.

As demonstrated in Section \ref{sec_motivation_data_collection}, about 73\% apps contain UI pages which require text input to pass. 
Providing appropriate text inputs during app testing remains a prominent obstacle that hinders access to app activities.
They involve 5 main input types, including \textit{identity}, \textit{geography}, \textit{numeric}, \textit{query} and \textit{comment}, and internal correlation between multiple inputs on one page.
Considering the diversity of inputs and contexts, it would require significant efforts to manually build a complete set of rules to deal with different input widgets.
These findings confirm the popularity of text input in mobile apps and the complexity of it for mobile testing, which motivates us to automatically generate meaningful text inputs from an existing knowledge base.

% and severity of text input for mobile testing and 

\subsection{Background of Large Language Model}
\label{sec_motivation_Large Language Model}
The target of this work is to generate the input text, and the Large Language Model (LLM) trained on ultra-large-scale corpus can understand the input prompts (sentences with prepending instructions or a few examples) and generate reasonable text.
% \chen{Not input query, but prompt, please explain what are prompts?} 
When pre-trained on billions of samples from the Internet, recent LLMs (like BERT~\cite{devlin2018bert}, GPT-3~\cite{brown2020GPT3} and T5~\cite{DBLP:journals/jmlr/RaffelSRLNMZLL20}) encode enough information to support many natural language processing tasks~\cite{lucy2021gender,sharples2022automated, yang2022empirical}. To invoke the desired functionality, users write natural language prompts~\cite{wang2022promda,gu2021ppt,liu2022p} as the LLM input, that are appropriate for the task.
% To boost the performance of LLM on different scenario, users often build prompts and answers for model prompt-tuning.

\begin{figure}[htb]
\centering
\vspace{-0.1in}
\includegraphics[width=8.3cm]{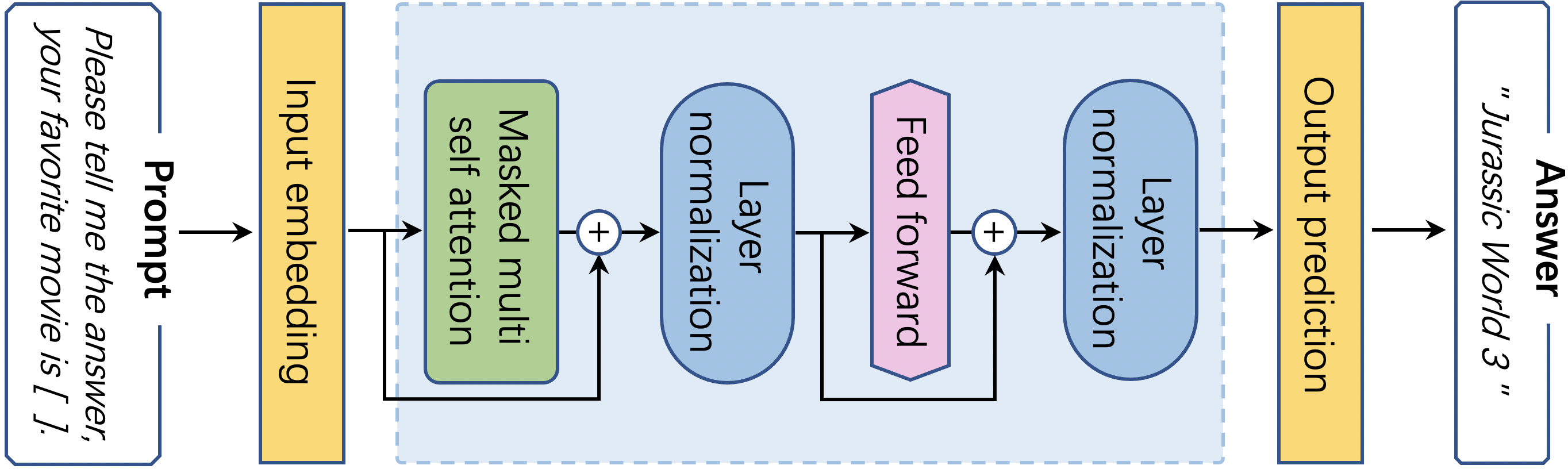}
\vspace{-0.05in}
\caption{The model structure of GPT-3.}
\label{fig:GPT-3}
\vspace{-0.05in}
\end{figure}

GPT-3~\cite{brown2020GPT3} is one of the most popular and powerful LLM which has great performance in many text generation tasks. GPT-3 is based on the transformer model~\cite{attention} including input embedding layers, masked multi-self attention, normalizaiton layers, and feed-forward in Fig \ref{fig:GPT-3}. 
% \chen{Any space to add some figure? Or some easy-to-understand examples from official OpenAI website.}
Given a sentence, the input embedding layer encodes it through word embedding. The multi-self attention layer is used to divide a whole high-dimensional space into several different subspaces to calculate the similarity. The normalization layer is implemented through a normalization step that fixes the mean and variance of each layer's inputs. The feed-forward layer compiles the data extracted by previous layers to form the final output. 

\begin{figure*}[t]
\centering
\vspace{-0.08in}
\includegraphics[width=18.0cm]{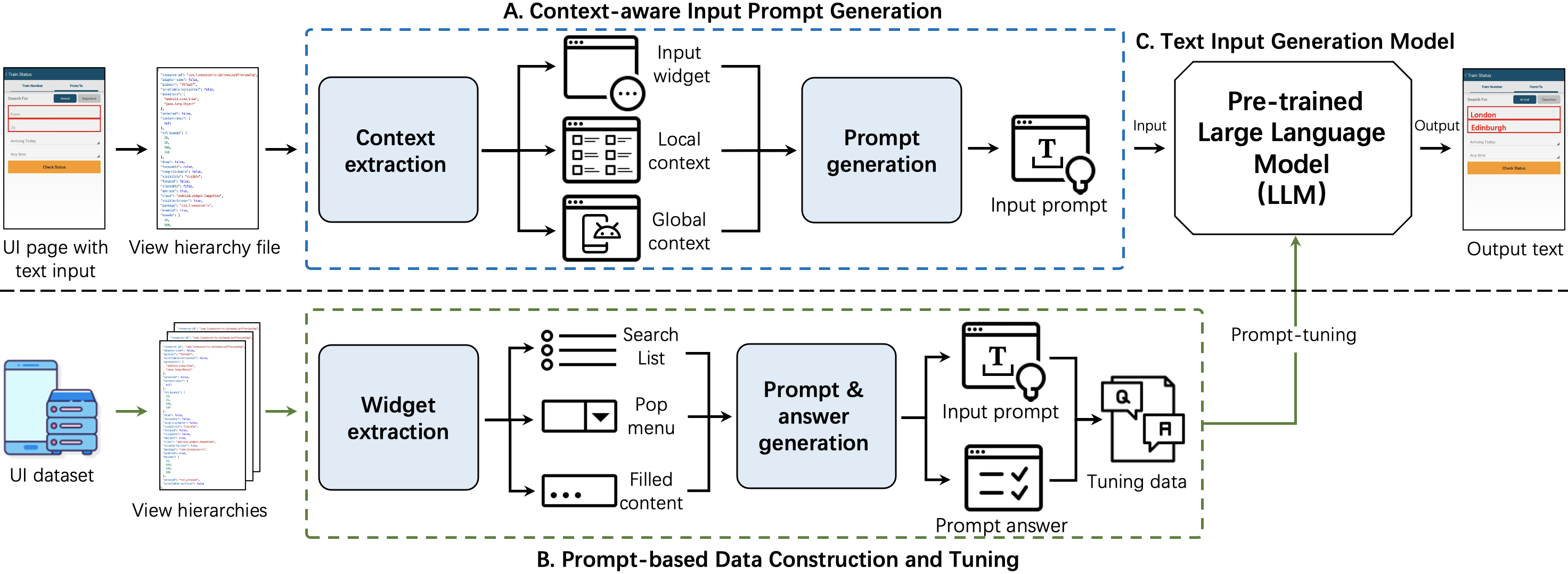}
% \vspace{-0.05in}
\caption{Overview of {\tool}.} 
\label{fig:overview}
\vspace{-0.15in}
\end{figure*}

\section{Approach}
\label{sec_approach}
This paper proposes a context-aware automated text input generation approach {\tool} to enhance mobile GUI testing. 
We employ the pre-trained Large Language Model (LLM) to intelligently generate semantic input text, as shown in Fig \ref{fig:overview}.
Specifically, given a GUI page with text inputs and its corresponding view hierarchy file, we first extract the context information for the text input, and design linguistic patterns to generate prompts as input for LLM.
To boost the performance of LLM on text input in mobile GUI, we develop a prompt-based data construction and tuning method to automatically extract the prompts and answers for model tuning. 

\subsection{Context-aware Input Prompt Generation}
\label{subsec_approach_Prompt_Generation}
Being trained on massive and diverse corpora as a headstart, LLM has demonstrated its outstanding performance on various tasks. 
Previous studies also reveal that its performance can be significantly influenced by the quality of its input, i.e., whether the input can precisely describe what to ask~\cite{chen2022knowprompt,liao2022ptau,zhou2022learning}.
Meanwhile, we observe that when using the app, rather than focusing on a certain text input widget, people tend to glance over the whole GUI page to determine the input content.
Motivated by this, we develop a context-aware input prompt generation method to construct the input prompt for LLM, in order to precisely describe the input widget information and facilitate the LLM outputting the desired answer.
We first extract the context information for the input widget, and design linguistic patterns for prompt generation.

\subsubsection{\textbf{Context Extraction}}
\label{subsubsec_approach_Information_Extraction}

As shown in Fig \ref{fig:overview}, for an input widget and its GUI page, we extract the context information from the view hierarchy file, which is easily obtained by automated GUI testing tools~\cite{machiry2013dynodroid,mao2016sapienz,gu2019practical,hao2014puma,DroidBot,su2017guided}. 
We extract the following three types of information in detail:

\textbf{Input widget information} is the most important source which describes inherent meaning of the input widget. 
The candidate information includes the ``hint-text'' field, ``resource-id'' field, ``text'' field of the input widget, and we choose the first non-empty one in order.

\textbf{Local context information} denotes the information nearby the input widget which facilitate the understanding of the context semantics. 
The candidate information source includes the parent node widgets, the leaf node widget, widgets with the same horizontal axis, and fragment of the current GUI page. 
For each information source, we extract the ``text'' or ``resource-id'' field (the first non-empty one in order), and combine this information together with the separator.

\textbf{Global context information} denotes the information related to high-level semantics of the input widget, which can further help refine the understanding of the input widget.
The candidate information source includes the current activity name, app name, and a number of input widgets on the GUI page. 
%We extract the ``activity\_name'' field, and will parse it to obtain the detailed information (Section \ref{subsubsec_approach_Prompt_Generation}). 
% \jie{comment}
% %这里没太懂，从activity name filed那里，就能获取到activity name和app name吗
% \liuzhe{comment}
% 这里是可以的，我之前给了一个例子，就是activity name这个字段前半部分是app name，后面是activity name
% 例如下面这个例子
% Property usually follows the naming rules suggested by Google, for example: ``com.company.appname.XXActivity''. We tokenize it according to `.' and extract the app name and activity name. 
% Then we traverse the view hierarchy file to get the number of input widgets on the current UI interface. 

% Global context information can help the model determine the type of generated text. We extract the current activity name, the input widgets number and the input content category as the app related information.
% For the app name and activity name, We extract the property corresponding to the ``activity\_name'' field in the view hierarchy file. 
% Then we traverse the view hierarchy file to get the number of input widgets on the current UI interface. 

% \input{tab/approach-rule}
\begin{table*}
\vspace{-0.1in}
\renewcommand\arraystretch{0.95} 
\caption{The example of linguistic patterns of prompts and prompt generation rules. }
% \jie{how about we add a note about the [n], [v+n], [prep], etc?}} 
% \chen{``related to'', not with}}
\vspace{-0.1in}
\label{tab:approach-rule}
\centering
\footnotesize
\begin{center}
\begin{tabular}{m{0.3cm}<{\centering} | m{7.1cm}<{} | m{9.25cm}<{}}
% \begin{tabular}{p{0.3cm}<{\centering} | p{8.0cm}<{\centering} | p{8.0cm}<{\centering}}
\toprule
\textbf{Id} & \textbf{Sample of linguistic patterns/rules} & \textbf{Examples of linguistic patterns/rules
instantiation}\\
\midrule
\multicolumn{3}{c}{\textbf{Patterns related to input widget: \hlr{$\langle IWPtn\rangle $}}}\\ \rowcolor{red!10}
\midrule
1 & Please input $\langle widget[n]\rangle $, the $\langle widget[n]\rangle $ is & Please input game name, the game name is  \\ \rowcolor{red!10}
2 & Please $\langle widget[v+n]\rangle $, the $\langle widget[n]\rangle $ is & Please search the food, the food is \\ \rowcolor{red!10}
3 & $\langle widget[n]\rangle $ + $[MASK]$ + $\langle widget[n]\rangle $ & Your weight is [MASK] kg\\ \rowcolor{red!10}
4 & $\langle widget[prep]\rangle $ + $[MASK]$ & From [MASK] \\ 
\midrule
\multicolumn{3}{c}{\textbf{Patterns related to local context: \hlb{$\langle LCPtn\rangle $}}}\\ \rowcolor{blue!10}
\midrule
5 & This input is about $\langle local[n]\rangle $ & This input is about the NBA team. \\ \rowcolor{blue!10}
% 6 & This input is about $\langle local[n]\rangle $, please $\langle local[v]\rangle $ & This input is about your health, please input. \\
6 & This input is about $\langle local[n]\rangle $, we need to $\langle local[v+n]\rangle $ & This input is about one-way flight, we need to search the flight information. \\ 
\midrule
\multicolumn{3}{c}{\textbf{Patterns related to global context: \hlg{$\langle GCPtn\rangle $}}}\\  \rowcolor{green!10}
\midrule
7 & This is $\langle app\ name\rangle $ app, in its $\langle activity\ name\rangle $ page, the input category is $\langle input\ category\rangle $. & This is a NBA sport app, in its search the NBA team page, the input category is query category. \\ 
\bottomrule
\toprule
\multicolumn{3}{c}{\textbf{Prompt generation rules}}\\
\midrule
1 & \colorbox{green!10}{$\langle GCPtn\rangle$} + \colorbox{blue!10}{$\langle LCPtn\rangle$} + \colorbox{red!10}{$\langle IWPtn\rangle$} & \hlg{This is a my movie app, in its search movie page, the input category is query category. } \hlb{ This input is about your favorite move in this year.} \hlr{ Please search the movie, the movie is } \\
\midrule
2 & \colorbox{green!10}{$\langle GCPtn\rangle$} + $[$\colorbox{blue!10}{$\langle LCPtn\rangle$} + \colorbox{red!10}{$\langle IWPtn\rangle$}$]$\{n\} & \hlg{This is a money wallet app, in its personal income page, the input category is numeric category.} \hlb{ This input is about your monthly income.} \hlr{ Income is [MASK] dollar.} \hlb{ This input is about your expenses.} \hlr{ Expenses is [MASK] dollar.}\\
% 9 & $\langle GCIP\rangle $ + $\langle LCIP\rangle $ + $\langle QIP\rangle $(``$\langle query[n]\rangle $'') & \hlm{This is my movie app, in its search movie page, the input category is query category. This input is about your favorite move in this year. Please search the movie, the movie is} \\
% \midrule
% 10 & $\langle GCIP\rangle $ + $\langle LCIP\rangle $ + $\langle QIP\rangle $(``$[MASK]$'') + $\langle LCIP\rangle $ + $\langle QIP\rangle $(``$[MASK]$''), ... & This is money wallet app, in its personal income page, the input category is numeric category. This input is about your monthly income and expenses. Income [MASK] dollar, expenses [MASK] dollar.\\
% \hline
\bottomrule
\end{tabular}
\end{center}
\vspace{-0.05in}
\begin{tablenotes}
\footnotesize
\item \textbf{\textit{Notes:}} ``[n]'', ``[v]'', ``[v+n]'' and ``[prep]'' means noun, verb, verb+noun and preposition in the related information. 
% \rev{of patterns related to input widget and local context.}
\end{tablenotes}
% \vspace{-0.15in}
\end{table*}

\begin{figure*}[t]
\centering
\vspace{0.02in}
\includegraphics[width=17.6cm]{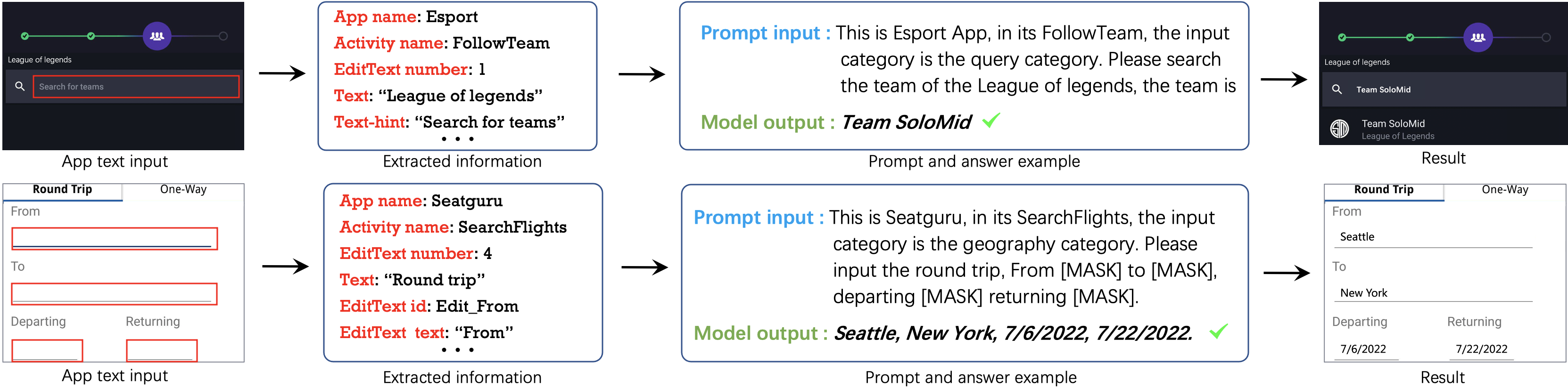}
\vspace{-0.04in}
\caption{Example of input prompt generation.} 
\label{fig:Example-input-prompt}
\vspace{-0.15in}
\end{figure*}

\subsubsection{\textbf{Prompt Generation}}
\label{subsubsec_approach_Prompt_Generation}
With the extracted information, we conduct preprocessing, and design linguistic patterns to generate prompts for inputting into the LLM.
% We integrate the above information into a prompt sentence with semantic information as the input of the model. Specifically, we first preprocess the sentences or phrases obtained from above three information, extract the prompt keywords and determine the input categories. Then we summarize three patterns for generating corresponding prompts. Finally, we generate prompt statements of input components through predefined rules. 

\textbf{Preprocessing:} 
We first tokenize the extracted information by the underscore and Camel Case~\cite{pascalcase} (e.g. Capital letters of each word) considering the naming convention in app development, and remove the stop words to reduce noise.
We then conduct the part-of-speech (POS) tagging of the information, so that the keywords can be highlighted. 
Specifically, we use the Standford NLP parser~\cite{de2008stanford} for that, and the noun, verb, prepositions are paid special attention to when designing the linguistic patterns. 

As demonstrated in Section \ref{sec_motivation}, apps exert diversified input contents which could influence the LLM to understand what should be output, thus when designing the linguistic patterns, we would take the input category into account.
We employ a straightforward keyword matching method for deriving the input category. 
Based on all 2,600 apps labeled in our pilot study, we summarize 5 glossaries (e.g., location, search, etc.)
% \chen{Do not understand.} 
corresponding to the 5 input categories. 
We then utilize the information in the activity name and app name, and check whether there is a match with one of the glossaries, following the category order in Section \ref{sec_motivation_types_UIs}. 

\textbf{Linguistic patterns of prompt:} 
To design the patterns, four authors write a prompt sentence for the input widget, input it into LLM for generating the input content, and check whether the content can make the app transition to a new activity. 
Each author can access to a random-chosen 500 apps from the pilot study, and he/she can obtain preprocessed information related with the input widget, the local context, and the global context. 
After 5 hours of trial, he/she is required to provide the most promising and diversified 30 prompt sentences, which are served as the seeds for designing patterns. 
With the prompt sentences, the four authors then conduct card sorting~\cite{spencer2009card} and discussion to derive the linguistic patterns as shown below. 
% \chen{Do we need this paragraph? Why not just directly tell these patterns?}

% \rev{we randomly choose 500 apps from pilot study and determine the text corresponding to the text input to evaluate whether the text generated by the pre-trained model is correct. 
% The four authors manually write the prompt sentence for the text input of these 500 apps according to the three aspects of app information, contextual widget and query widget. Finally, they input the prompt sentence into the pre-trained model to determine whether the correct text can be output (enter the output and app can pass successfully). They choose the correct prompt sentences and summarize the prompt patterns and generation rules through discussion.}

% To identify the patterns, we select 500 apps from pilot study and determine the text corresponding to the text input to evaluate whether the text generated by the pre-trained model is correct. The four authors manually write the prompt pattern for the text input of these 500 apps according to the three aspects of app information, contextual widget and query widget. Then they generate the final prompt sentence according to the generation rules. Finally, they input the prompt into the pre-trained model to determine whether the correct text can be output. They combine and summarize the correct prompt sentences to get the corresponding pattern. They discuss the problematic prompt and get the final pattern.

This process comes out with 14 linguistic patterns respectively related to input widget, local context and global context; and Table \ref{tab:approach-rule} shows 8 of them.
% \chen{Why only show 8?}
The patterns of the input widget explicitly specify what should be input into the widget, and we employ the keywords like noun ($widget[n]$), verb ($widget[v]$) and preposition ($widget[prep]$) for designing the pattern. 
% \rev{, and extract noun ($widget[n]$), verb ($widget[v]$) and preposition ($widget[prep]$) from input widget as keywords.}
We design two forms of expression, i.e., continuation (e.g., ``the game name is:'') and mask (e.g., ``your weight is [MASK] kg''), to accurately depict the characteristics of the input widget. 
The patterns about the local and global context are used to enhance the prompt from the viewpoints of whole GUI page and whole app. For example, the global pattern encodes the information about the app name, activity name, and input category.
% \rev{$\langle local[v+n]\rangle$ shows ``search the flight information''.}
% \jie{comment}
%这个pattern看起来和第一类没啥区别
% \jie{comment}
%表格里面，不要用IP吧，这个感觉是知识产权的缩写，外国人看起来容易跳戏。就叫做QPtn？ pattern用辅音字符代替。然后三个名字改一下， patterns related with  query widget， local context ， global context 吧。

% \rev{We summary 3 categories and 14 Linguistic patterns to construct the prompt information of the text input. Table  shows the sample of sub-patterns, detailed information can be find in our GitHub. (1) Query information pattern ($\langle  QIP\rangle $): The query information of text input is the main part of the prompt, we design two forms of expression ``($query[n]$)'' and ``[MASK]''  for different scenarios. (2) Local context information pattern ($\langle  LCIP\rangle $): The local context information pattern of text input is used to enhance the semantics of the prompt. (3) Global context information pattern ($\langle  GCIP\rangle $): The app information pattern of text input is used to enhance the input category and the function of the input UI page. }

\textbf{Prompt generation rules:} 
Since the designed patterns depict the information from different points of view, we then combine the patterns from different viewpoints together and produce the prompt rules as shown in Table \ref{tab:approach-rule}. 
% \chen{May need to briefly introduce these 3 kinds of patterns?}
There are two rules respectively for two typical cases, i.e., GUI page with one input widget and GUI page with multiple input widgets. 
For the former case, we simply combine the patterns from the three sources of information together as Rule 1 in Table \ref{tab:approach-rule}. 
And for the latter case, since all input widgets in a GUI page share the same global context, therefore we first utilize the pattern about the global context, followed by the pattern with local context and input widget for each text input as Rule 2 in Table \ref{tab:approach-rule}. 
Due to the robustness of the LLM, the generated prompt sentence does not need to follow the grammar completely.
We also provide examples in Fig \ref{fig:Example-input-prompt} to illustrate how the rules work.

\subsection{Prompt-based Data Construction and Tuning}
\label{subsec_approach_Prompt-tuning}

The general pre-trained model can hardly perform well on domain-specific tasks (e.g., app text input generation), one common practice is to prompt-tune~\cite{chen2022knowprompt,zhou2022learning,wu2022fast,wang2022promda,cui2022prototypical,gu2021ppt,liu2022p} the model to let it understand and recognize the input prompt syntactic~\cite{DBLP:journals/corr/abs-2207-11680,liao2022ptau,chen2022knowprompt,zhou2022learning}. 
However, there is so far no such type of open dataset with input prompts and corresponding answers.
Manually data collection is very time- and effort-consuming, and human bias and lack of diversity may further degrade the data quality. 
Meanwhile, we notice that some widgets are similar to input widgets, and have candidate or pre-input content (like answers), which naturally serve our purpose. 
%provides us the potential to automatically construct the dataset for prompt-tuning. 
Therefore, we develop an approach to automatically construct a prompt-based dataset from the view hierarchy files to prompt-tune the pre-trained model.

% \input{figure/4-kind-data}

% \jie{comment}
% %default or pre-input answers 这里除了answer 我们还能用别的词吗？感觉这里其实是输入的内容。answer这个词其实是用到我们这个模型里面才是用answer，但其实这里感觉还是应该从问题层面说，其实就是和input widget类似，同时又包含这个输入内容的。你和我后面某个批注一起看。
% %而且你再看看你这个图3哈，图和现在的文字对应不上，tuning data extraction就不太知道是个啥。我觉得这块还需要想想，这块我们要显示的分出两个小章节来吗？
% %要不这里tuning data extraction 就是要把 alternative input widget和input content明确的找出来？然后后面的prompt answer construction就是得到这个prompt和answer，那第二步感觉就没啥内容。因为我感觉组件和答案的对应关系应该也是第一步得到了。这块你连同我后面的批注 再想想。
% \liuzhe{comment}
% % 这块儿内容感觉一步就差不多了,找到这些组件,分出组件的prompt和answer
% % 或者我们叫成组件确认+prompt和answer构建这两步.第一个步骤就是描述一些这个组件的信

\begin{figure}[htb]
\centering
\vspace{-0.05in}
\includegraphics[width=8.6cm]{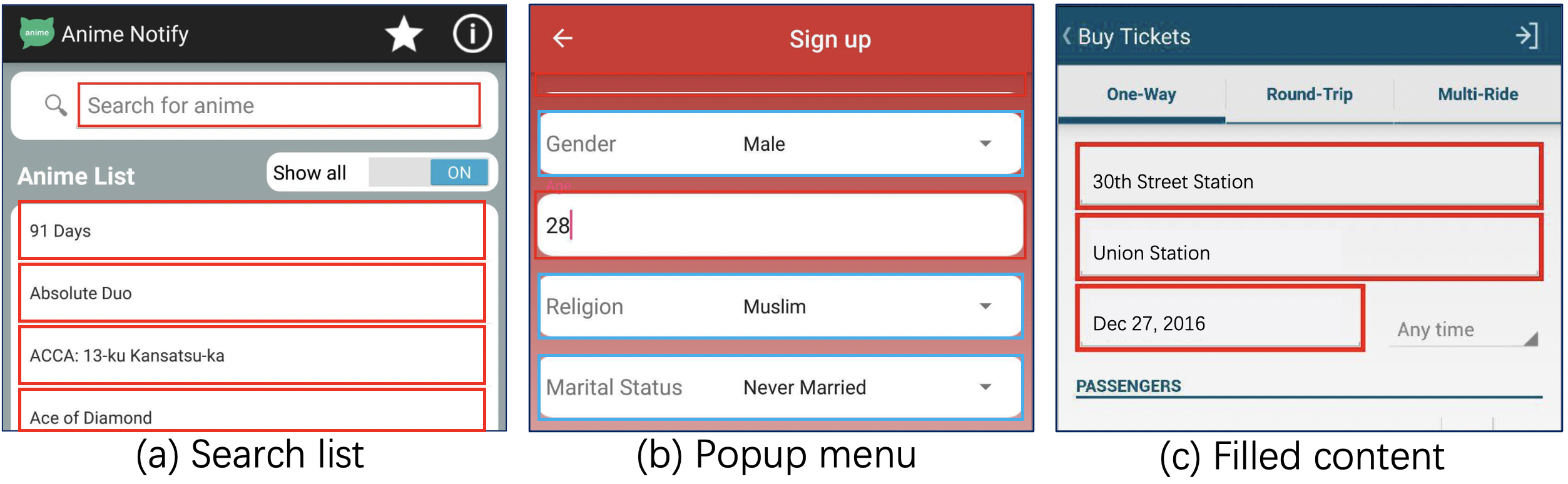}
% \vspace{-0.1in}
\caption{3 example cases of data construction.}
\label{fig:3-kind-data}
\vspace{-0.1in}
\end{figure}

Specifically, we find 3 cases that can be utilized for data construction as demonstrated in Fig \ref{fig:3-kind-data}.
(1) Some input widgets contain associated lists for the candidate input contents which can be treated as already having the input.
(2) Popup menu is another variant of the input widget, which has the candidate input content for data construction.
(3) Some open datasets with view hierarchy files are collected by crowd workers (Rico~\cite{Rico} is the largest repository of mobile app UIs collected by crowdsourcing),
% \chen{Too complicated, maybe just use the words ``hints''} 
so some input widgets contain the pre-filled content.
We regard these input widgets with candidate input contents as the explicit input widgets, and describe how we utilize these widgets for data construction. 

\subsubsection{\textbf{Widget Extraction}}
%\subsubsection{\textbf{Explicit Input and Related Content Retrieval}}
The challenges of constructing prompt and answer are to find the explicit input widget, and map its candidate input contents with the widget.  

% Then we use the method in Section \ref{subsec_approach_Prompt_Generation} to obtain the input prompt for the explicit input widget, and treat the candidate input contents as the answer. 
% We will introduce the construction method for the three cases. 
% \jie{comment}
%我这里调整了下，下面就是怎么找explicit input widget，以及对应的input content的。这里一句话讲了有了这些内容，我们是怎么做的。因为这块对于每类都是这样的，所以就在这里一起说了，这样逻辑比较好。

% \rev{from existing dataset is finding alternative input widget with similar function as text input and it has the answer. After localizing the above 3 alternative input widgets, we propose the prompt and answer construction algorithm to construct the prompt and answer through widget information and its content. First, we extract the prompt information and its corresponding answer information from 3 alternative input widgets. Based on the coordinates and size of them, the algorithm determines and adjusts the relationship between their positions according to specific rules to construct the prompt sentence ($prompt$) and its answer ($answer$). Please note that our the same prompt generation method in Section  to get the $prompt$ from prompt information. 
% We will introduce different types of answer construction methods as below:}

\textbf{Extraction from search list}: For this case, there is usually an input widget (\textit{EditText}) and its associated list (\textit{ListView}). 
We first check whether there is a \textit{ListView} directly below the \textit{EditText}, if it is true, we treat the \textit{EditText} as the explicit input widget.
For the candidate inputs contents, 
%we extract it from the \textit{ListView}.
if all items in the \textit{ListView} are \textit{TextView} (e.g., a word or phrase as shown in Fig \ref{fig:3-kind-data} (a)), the text is directly treated as the candidate input content. 
Otherwise, we only extract the title of the time as the candidate input content by checking the \textit{TextView} with the smallest ordinate value in each item, if it contains the title and body (e.g., the news list with title and publication time). 

\textbf{Extraction from popup menu}: 
Popup menu is triggered when clicking a \textit{Spinner}, and contain a list of \textit{TextView}.
As shown in Fig \ref{fig:3-kind-data} (b), we first find a \textit{Spinner} widget, and retrieve the \textit{TextView} whose abscissa is less than the minimum abscissa of the \textit{Spinner} or ordinate is less than the minimum ordinate of \textit{Spinner} in order.
We then treat the \textit{TextView} as the explicit input widget, and the ``hint-text'' field of \textit{Spinner} as the candidate input content.
% \jie{comment}
%这里我怎么觉得应该是横坐标相同，纵坐标比最小的小？感觉两个坐标都应该限制下。
% \liuzhe{comment}
% 这里我想表达这样一个意思，就是按照顺序找出（1）纵坐标相同，横坐标比较小的那个，（2）横坐标相同，纵坐标比较小的那个。按照顺序（1）有了就不管（2）了，不知道改制后的这个写法有没有歧义

% The popup menu (\textit{Spinner}) is a special input widget that its content can be used as an answer, and we need to extract it corresponding alternative input widgets as prompt. 
% As shown in Figure \ref{fig:3-kind-data} (b), the widgets on the left often have prompt information. We extract the \textit{TextView} whose abscissa is less than the minimum abscissa of the \textit{Spinner} and its text field is not ``none''. Then, we regard the ``hint-text'' field of \textit{Spinner} as $answer$ and the ``text'' in \textit{TextView} as prompt information.

% By analyzing whether the \textit{TextView} ($Tx_2$) whose abscissa is less than the minimum abscissa of the \textit{Spinner} ($Sx_1$) has text information, we regard the \textit{Spinner} as answer and the \textit{TextView} as prompt.
% \jie{comment}
% %这个has text information 具体标准是啥？text field不为空？
% \liuzhe{comment}
% % 嗯嗯是的，不为空

\textbf{Extraction from filled content}: The content filled by the crowd workers
% \chen{Mentioning crowd workers makes it very difficult to understand, may just say hints of the editText...} 
is displayed in the property of ``hint-text'' of \textit{EditText}, we filter the cases of preset content by the developers, and remain the ones filled by humans. 
This is done through filtering the \textit{EditText} whose ``hint-text'' contains ``search, add, input, enter, etc'' which is the preset information with high probability. 
We then treat the \textit{EditText} as explicit input widget, and use the ``hint-text'' field as input content for data construction. 
% \chen{try not to use ``will'', ``need to/should'', but just state steps with the present tense in the approach}

% \jie{comment}
%是包含，还是什么。如果众测人员输入的一长串内容，也恰好包含了某个关键字呢？
% \liuzhe{comment}
% 包含其中的一个就不算
% 如果众测人员输入长内容且包含的话，我们这里也不算，就是模棱两可的都不算

\subsubsection{\textbf{Prompt and Answer Generation}}
After obtaining three types of explicit input widgets and their candidate input contents, we generate pairs of prompt and answer for model tuning. 
Specifically, we use the method in Section \ref{subsec_approach_Prompt_Generation} to obtain the input prompt for the explicit input widget, i.e., input the explicit input widget and its context information to get the input prompt sentence. 
And we treat the candidate input contents as the answer. 
% We then build the input prompt and prompt answer into our tuning data. 

We then fine-tune the model with the prompt as the input and the answer as the output (the same learning objective as the pre-trained model).
This prompt-tuning stimulates pre-trained model to recognize the input prompt syntactic and usage patterns of the fine-tuning data~\cite{zhou2022learning,wu2022fast,wang2022promda,cui2022prototypical,gu2021ppt,liu2022p}.
% , rather than simply memorizing some text input and its corresponding content.
% \chen{Do not understand}
% \liuzhe{comment}
%这里我想表达的是prompt tuning的做法是让模型了解我们这种prompt输入的模式，输出是什么样的内容，例如：[mask]类型要填写xxx
% 这里要不我引用一下，这个说法是做prompt tuning常用的做法
% \jie{comment}
%这句话我稍微改了一下，但感觉还是不太通顺，特别是rather than那半句。你看看已有文章里面是怎么说这个事情的，直接用一下
%我把说小数据集的这个事情去掉了，感觉是减分项。
%然后我把那个小标题去掉了。
Therefore, it can be more effective than general fine-tuning in training the model in understanding the questions.

% Then we use the method in Section \ref{subsec_approach_Prompt_Generation} to obtain the input prompt for the explicit input widget, and treat the candidate input contents as the answer. Specifically, we input the explicit input widget and its context information to get the input prompt sentence, and then build input prompt and prompt answer into our tuning data.

% \subsubsection{\textbf{Prompt-tuning}}

% Through the above algorithm, we obtained the tuning dataset including 7,000 pairs of prompt and answer from Rico~\cite{Rico} and FrontMatter, which is used to prompt tune the pre-trained model. 
% Specifically, 
% We fine tune the model with the prompt as the input and the answer as the output (the same learning objective as the pre-trained model). We only need a small amount of data to achieve highly effective prompt-tuning. This is because prompt-tuning stimulates pre-trained model to recognize the input prompt syntactic and usage pattens in the large language data which it is trained, rather than simply memorizing some input text and limited usage contexts used for model fine tuning.

\subsection{Text Input Generation Model}
\label{subsec_approach_Generation_Model}
% \jie{comment}
%这一部分压缩一下吧。

The core part of {\tool} is the pre-trained Large Language Model (LLM), which has been proven to be effective in many downstream tasks, e.g., question answering, article generation, etc.~\cite{laskin2020reinforcement,yang2022empirical,chen2020big}.
Several LLM has been proposed and utilized in various fields, e.g., GPT-3, BERT, RoBERTa, T5.
This paper chooses GPT-3~\cite{brown2020GPT3} (Generative Pre-trained Transformer-3), which is the state-of-the-art LLM from OpenAI with 175 billion parameters trained on billions of tokens that can smartly generate semantic text. We use the input prompt obtained in Section \ref{subsec_approach_Prompt_Generation} as the input of the model and feed it into the GPT-3 model. The model output is used as input content to enhance automated testing.
The model details are shown in Section \ref{sec_motivation_Large Language Model}.

\subsection{Implementation}
\label{subsec_approach_Implement}

% \jie{comment}
%这里我们要明确，我们是在C里面说用了GPT-3还是在这里说。需要两个地方都说吗？或者在上面说了，这里是需要说我们用openAI提供的模型，大概说一下模型的细节。而不应该在说我们select了什么，这个select已经在前面说了
We implement our text input generation model based on the pre-trained GPT-3 model which was released on the OpenAI website
\footnote{\url{https://beta.openai.com/docs/models/gpt-3}}. 
% Our large language model selects the most popular and powerful model~\cite{brown2020GPT3}, Generative Pre-Train Model (GPT-3) provided by OpenAI, which is widely used in a large number of scenes. 
% GPT-3 has 175 billion parameters and can adapt to more scenarios through prompt tuning. 
The basic model of GPT-3 is the Curie model which is extremely powerful and good at answering questions, the OpenAI CLI version is 0.9.4, the batch size is 64, the epoch is 100, and the learning rate multiplier is 0.01.
% \chen{Tell how long does it take to fine-tune}
We construct the tuning dataset including 7,000 pairs of prompts and answers from Rico~\cite{Rico} and FrontMatter~\cite{kuznetsov2021frontmatter}, which is used to prompt tune the GPT-3 model, and the prompt-tuning time is about 12 hours.
% \jie{comment}
%这句话是不是应该往上面放，先把模型相关的都说完了。再说tuning dataset那个。你现在逻辑有点乱吧
{\tool} obtains the view hierarchy file of the current UI page through UIAutomator~\cite{uiautomator} to analyze and extract the text input widget information.
{\tool} can be integrated by replacing the text input generation module of the automated GUI testing tool, which automatically gets the input widget in the view hierarchy file and generates its corresponding text to improve activity coverage.

\section{Effectiveness Evaluation}
\label{sec_Effectiveness}

In order to verify the text generation performance of {\tool}, we evaluate it by judging whether the generated text input can pass the UI page requiring text input.
We will further evaluate the usefulness of {\tool} in real-world practice by integrating it with automated GUI testing tools in the next section.

% \jie{comment}
%这块我觉得前面加几句，这里是怎么evaluate的，就是看某个UI page 它能不能通过，然后说在section XX，我们还会验证将该方法集成到automated testing tools之后的实际应用效果。这样能够让人有个总体的认知。
%RQ2 要不就直接叫 what is the quality of the XX ? 感觉how accurate 不太全面

% \jie{comment}
%在题目里面说passing rate 大家不知道是要说啥，所以要明确写出来。
%这里你看看RNN那个工作是怎么写的。我不知道making the app pass current UI page 这个说法是否地道
%在题目里面说passing rate 大家不知道是要说啥} 

\begin{itemize}[leftmargin=*]
\item \textbf{RQ1: How effective is our proposed {\tool} in passing UI pages requiring text inputs?}
% \chen{If agree, please update the whole section.}

\end{itemize}
For RQ1, we compare the passing rate of {\tool} compared with 11 commonly-used and state-of-the-art approaches (details are in Section \ref{subsec_experiment_baseline}).

\begin{itemize}[leftmargin=*]
\item \textbf{RQ2: What is the quality of the input text generated by {\tool}?}
\end{itemize}
For RQ2, we ask 20 testers to manually check the quality of its generated input text.

\subsection{Experimental Setup}
\label{subsec_experiment_dataset}

% \jie{comment}
%这句话不清晰，most popular 637 apps，这个updated since May 2022 起到的作用是什么？popular是根据updated的时间来看的吗？不是吧？所以既popular又得在这个时间点之后更新过？
%Droidbot第一个字母要不要大写，统一
For RQ1, we crawl the 100 most popular apps of each category from Google Play~\cite{Googleplay}, and only keep the latest ones with at least one update after Mar. 2022, resulting in 637 apps in 12 app categories. %in Google play, this comes out with 637 apps. 
Then, we use DroidBot to automatically run these apps to explore the app for locating GUIs requiring text inputs by analyzing run-time view hierarchy files.
Note that we filter out some apps by the following criteria: (1) UIAutomator~\cite{uiautomator} can't obtain the view hierarchy file; (2) they would constantly crash on the emulator; (3) one or more baselines can't run on them; (4) they have no pages requiring text inputs. 
Finally, 106 apps with 168 text inputs remain for further experiments. 
% one or more baselines cannot run it; (2) 
% to analyze the text input number. 
% Apps are filtered as follows: (1) All baseline tools can run them. (2) They do not constantly crash on our emulator, (3) UIAutomator~\cite{uiautomator} can get the view hierarchy file of UI page. (4) Each app has text inputs. Finally, we obtained 100 apps with text inputs for experiments.
% because the existing researches do not provide apps with text input (some apps have been taken off the shelves or do not provide download APK~\cite{liu2017automatic})
% \jie{comment}
%前面这句话就没太有必要吧。就说我们选了pupular apps，
%according to the update time (after January 2022)，这个是怎么选的，2022年1月之后有更新的app里面，随机选择了100个？说法不太明确。
% and various categories of our pilot study~\cite{he2020textexerciser},
% we randomly choose 100 popular mobile apps according to text inputs (each app has 1 to 5 inputs), input category and update time (after January 2022) from Google Play.
% \footnote{http://play.google.com/store/apps}
% , each app has at least one page containing 1 to 5 text inputs. 
% \jie{comment}
%has at least one page? 我觉得你就是每个app random 选择了一个至少带有一个输入的page，然后最后每个page上面包含1-5个inputs。
%我看了后面知道你要说哈了，那这里就是说每个app至少包含1个输入就行了。后面才是随机选的
% \liuzhe{The apps that we choose need to work properly on our testing infrastructure: they do not constantly crash on our emulators.}
The details of the apps are shown in Table \ref{tab:app-info}. We employ passing rate (passing UI pages requiring text inputs), which is a widely used metric for evaluating GUI testing~\cite{he2020textexerciser,liu2017automatic,arnatovich2018mobolic}.
Specifically, we randomly select one UI page with text input from each app and get the view hierarchy file through UIAutomator, input the text generated by {\tool}
% \jie{comment}
%这个input text是模型产生的input？那是不是还得说，对于某个页面，是怎么拿到view file的？
%if the inputs are prompted with an error or do not get the expected result after entering the text，这句话是什么？模型没有产生结果吗？我觉得不就是把模型产生的结果输入进去，看有没有pass吗？
%有没有pass这个事情很难界定吗？需要5个tester来判断？还会有disagree？
% \liuzhe{comment}
% 这里想说如果输入的文本输入app之后提示输入错误，或者没有达到预期效果的，我们就判定是没有通过
% 感觉好像不用说的这么复杂
and check whether it passes UI pages requiring text inputs.
% \rev{the correct text input passing.}
% Please note that if the inputs are prompted with an error or do not get the expected result after entering the text, it is considered that the transition is not pass. 
% This process has been verified by 2 testers with more than 5 years of testing experience.  \chen{Why do we need humans to check?}
% \liuzhe{comment}
% 这里感觉也可以纯自动化，不知道要不要说一下怎么判断？
% After each state transition, they will compare the evaluation results and if there is any disagreement, they will discuss until the consensus is reached.
To further evaluate the advantage of {\tool}, we additionally compare it with 11 common-used and state-of-the-art baselines (details are in Section \ref{subsec_experiment_baseline}).
We design the script for each baseline to ensure that it can cover the text input page, and each tool is repeated three times in the same experimental environment (Android x64 emulator) to mitigate potential bias; as long as one pass occurs, we treat it as successful passing.
% \rev{ all times and inputs passing are recorded as successful passing.}
% For the automated GUI testing tools, we extracted its input text generation module for experiment.
% \jie{comment}
%这句话感觉有风险，可能会被问，你是怎么提取的？要不就不要了？没法说我们只让工具执行一个页面是吧？
%table 2 我稍微调整了下

\begin{table}[h]
\vspace{-0.05in}
\renewcommand\arraystretch{0.95} 
\caption{Dataset of effectiveness evaluation.}
\vspace{-0.05in}
\label{tab:app-info}
\centering
\footnotesize
\begin{tabular}{p{2.6cm}<{\centering} | p{2.9cm}<{\centering} | p{1.9cm}<{\centering}}
\toprule
\textbf{Input category} & \textbf{\# apps (UI pages)} & \textbf{\# inputs}\\ 
\midrule
Identity & 20 & 36 \\
Geography & 20 & 34 \\
Numeric & 20 & 33\\
Query & 26 & 39 \\
Comment & 20 & 26 \\
\midrule
Overall & 106 & 168 \\
% \hline
\bottomrule
\end{tabular}
% \vspace{0.05in}
% \begin{tablenotes}
% \footnotesize
% % \item \textbf{\textit{Notes:}} ``App number'' is the number of apps, each app we choose one GUI page with text inputs.
% \end{tablenotes}
\vspace{-0.05in}
\end{table}

For RQ2, we conduct a user study to verify the quality of {\tool}'s generated input text.
We recruit 20 software testers online, all of whom major in computer science and have more than 3 years of software testing experience.
Each of them is presented with the UI page of the app, the input text generated from {\tool} and baselines in RQ1.
They are required to independently evaluate the input text generation results, and to answer the question whether they agree with each of the generation results using a 5-Likert scale, i.e., strongly agree, agree, neutral, disagree, strongly disagree~\cite{5-likert-1,brooke1996sus}.
We employ Kendall's W (Kendall's coefficient of concordance)~\cite{Kendall} to assess the agreement of the results among different practitioners. 
The closer the test outcome is to 1, the higher agreement among the evaluation results of the raters.
The evaluation results should be returned within 8 hours to ensure the credibility of this study.

\subsection{Baselines}
\label{subsec_experiment_baseline}
To demonstrate the advantage of {\tool}, we compare it with 11 common-used and state-of-the-art baselines. 
We roughly divide them into random-/rule-based methods, constraint-based methods, and learning-based methods, to facilitate understanding.
The 8 random-/rule-based methods are from automated GUI testing tools, and the other 3 methods are designed specifically for app input generation. 

For random-/rule-based methods, we use Stoat~\cite{su2017guided}, Droidbot~\cite{li2017droidbot}, Ape~\cite{gu2019practical}, Fastbot~\cite{cai2020fastbot}, ComboDroid~\cite{wang2020combodroid}, TimeMachine~\cite{dong2020time}, Humanoid~\cite{li2019humanoid}, Q-testing~\cite{pan2020reinforcement}. 
For constraint-based methods, we use Mobolic~\cite{arnatovich2018mobolic} and TextExerciser~\cite{he2020textexerciser}. 
% Mobolic generated and executed by combining the online testing technique and customated constraint input generation. \chen{Do not understand.}
Mobolic generated the input text by combining the online testing technique and the input constraint. 
% \jie{comment}
%这两个顺序我换了下。但感觉Mobolic，我没看明白是怎么做的。按照应该重点说inputs是怎么生成的。
TextExerciser's key idea is that Android apps often provide feedback, called hints, for malformed inputs, so the approach utilizes such hints to improve the input generation. 
For the learning-based method, Liu et al. ~\cite{liu2017automatic} utilize the RNN model and Word2Vec to predict the text input value for a given input widget (denote it as RNNInput for simplicity). 

\subsection{Result}
\label{subsec_results}
\subsubsection{\textbf{Passing Rate by {\tool} (RQ1)}}
\label{sec_results_RQ1}

Table \ref{tab:RQ1-result} shows the passing rate of {\tool} and the baselines.
{\tool} achieves the passing rate of 0.87 across 106 UI pages.
% \jie{comment}
%这里不应该说100 mobile apps，容易引起误解，其实是100 UI page。 上面那个table II 要不也改成UI pages 这种？
This indicates the effectiveness of {\tool} in passing most of the transition so as to
% \rev{provide a full viewpoint in augmenting}
facilitate the automated GUI testing covering more pages. 
% \jie{comment}
%我感觉usefulness那里，更适合用augment。这里的话也没有和自动化工具集成。

\begin{table}[h]
\vspace{-0.05in}
\renewcommand\arraystretch{0.95} 
\caption{Result of passing rate. (RQ1)}
\vspace{-0.05in}
\label{tab:RQ1-result}
\centering
\footnotesize
\begin{tabular}{p{1.6cm}<{\centering} | p{0.8cm}<{\centering} | p{0.6cm}<{\centering} | p{0.6cm}<{\centering} | p{0.8cm}<{\centering} | p{0.8cm}<{\centering} || p{0.6cm}<{\centering}}
% \hline
\toprule
\textbf{Method} & \textbf{Ident} & \textbf{Geo} & \textbf{Num} & \textbf{Query} & \textbf{Comm} & \textbf{All}\\ 
% \hline
\midrule
\multicolumn{7}{c}{\textbf{Random-/rule-based method}}\\
% \hline
\midrule
Stoat & 0.10  & 0.10  & 0.05  & 0.15  & 0.60  & 0.20 \\ 
Droidbot & 0.10  & 0.05  & 0.00  & 0.15  & 0.60  & 0.18 \\ 
Ape & 0.20  & 0.15  & 0.10  & 0.12  & 0.65  & 0.24 \\ 
Fastbot & 0.15  & 0.10  & 0.05  & 0.12  & 0.65  & 0.21 \\ 
ComboDroid & 0.15  & 0.05  & 0.10  & 0.19  & 0.60  & 0.22 \\ 
TimeMachine  & 0.10  & 0.15  & 0.05  & 0.15  & 0.65  & 0.22 \\ 
Humanoid & 0.15  & 0.10  & 0.05  & 0.15  & 0.60  & 0.21 \\ 
Q-testing & 0.10  & 0.15  & 0.10  & 0.15  & 0.65  & 0.23 \\
% \hline
\midrule
\multicolumn{7}{c}{\textbf{Constraint-based method}}\\
% \hline
\midrule
Mobolic & 0.25 & 0.15 & 0.25 & 0.15 & 0.65 & 0.28 \\ 
TextExerciser & 0.45 & 0.15 & 0.40 & 0.23 & 0.70 & 0.38 \\ 
% \hline
\midrule
\multicolumn{7}{c}{\textbf{Learning-based method}}\\
% \hline
\midrule
RNNInput & 0.35 & 0.40 & 0.25 & 0.50 & 0.75 & 0.45 \\
% \hline
\midrule
% \multicolumn{7}{c}{\textbf{Our Method}}\\
% \hline
% QTypist (-PT) & 0.37  & 0.26  & 0.26  & 0.28  & 0.72  & 0.37 \\ 
QTypist (-T) & 0.55 & 0.65 & 0.50 & 0.58 & 0.80 & 0.61 \\ 
\textbf{QTypist} & \textbf{0.85}  & \textbf{0.90}  & \textbf{0.85}  & \textbf{0.85}  & \textbf{0.90}  & \textbf{0.87} \\
% \hline
\bottomrule
\end{tabular}
\vspace{0.05in}
\begin{tablenotes}
\footnotesize
\item \textbf{\textit{Notes:}} ``Ident'' is the identity information, ``Geo'' is the geographic information, ``Nume'' is the numerical information, ``Comm'' is the comment content. ``All'' is all data.
% QTypist (-PT) is QTypist without genenrate prompt and prompt-tuning. 
QTypist (-T) is QTypist without prompt-tuning.
\end{tablenotes}
\vspace{-0.1in}
\end{table}

{\tool} is 93\% (0.87 vs. 0.45) higher even compared with the best baseline (RNNInput).
% \jie{comment}
%这里我修改了下，我记得春阳老师好像之前提到过，这种地方他一般会从自己方法的优点这样的角度来说.
%现在就是比较精简，我感觉还不错，该表达的也表达了。这个per type这种事情，感觉非常的琐碎
Compared with RNNInput, our proposed approach takes full advantage of the context information around the text inputs, and employs the capability of the pre-trained Large Language Model for better understanding the input text.
% \rev{trains the model according to the type of app, and its robustness is affected by the type of app, so it is difficult to cover more kinds of apps. 
% For example, RNNInput only performs well in the query category text input generation tasks of movie and music, because this part of data appears in large numbers in the training data. When the training data is not updated in time or does not cover the input categories, the performance of the RNNInput will decline.}
% RNNInput is sensitive to movie and music categories for the input components of search content, and most of other categories are random input.
While other baseline methods mainly generate text randomly or with heuristic rules, resulting in the generated text being greatly affected by the template and with low robustness. 
For the constraint-based method as TextExerciser, since the feedback-related widgets only cover a small portion of all cases, e.g., most geographic-related input widgets would not involve the feedback, the related performance is poor. 
% \rev{For example, TextExerciser generates text through the constraints of input widgets. For unconditionally constrained text (search, geographic, etc.), its text generation is poor.}
We also observe that the baselines have a high passing rate in the comment category, mainly because texts without semantics entered in this category can pass. Therefore, we need to further evaluate the quality of generated text through RQ2.

% \jie{comment}
%我感觉第一句说的应该是random-rule based那些，第二句又说的是constraint那类。
%LLM是large language model，我看introduction是这么用的，所以这里就不要用large-scale 
%这里为啥不说a啊，只说b，c，那a放在这里起到啥作用了。

\begin{figure}[htb]
\centering
\vspace{-0.1in}
\includegraphics[width=8.6cm]{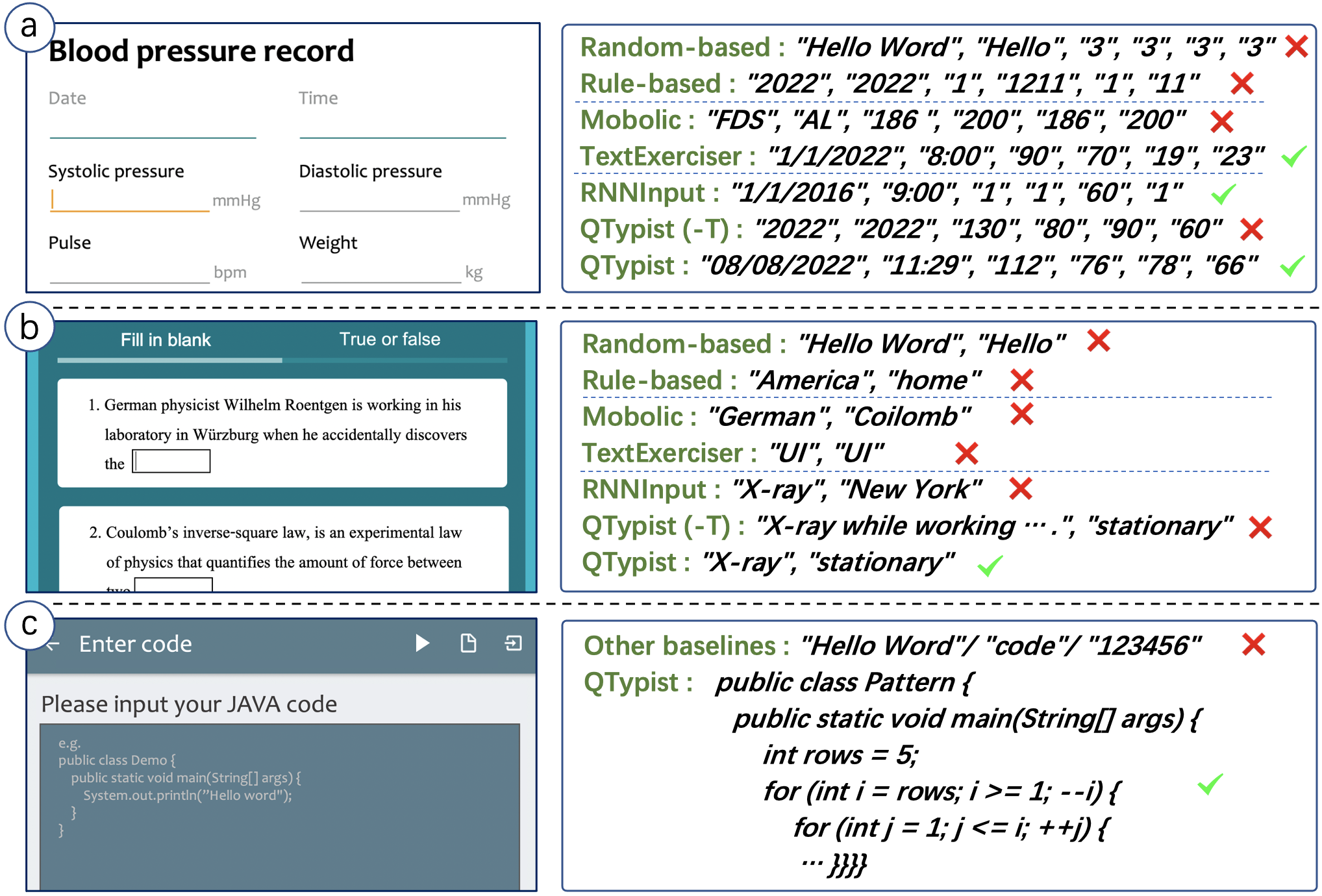}
% \vspace{-0.1in}
\caption{The example of input text of {\tool}.} 
% \chen{For (b), can ``hello world'' also pass it?}\liuzhe{cannot}}
\label{fig:RQ1-goodcase}
\vspace{-0.1in}
\end{figure}

Fig \ref{fig:RQ1-goodcase} demonstrates the advantages of {\tool} compared with baselines with illustrative examples. 
We can see that {\tool} performs better in more scenarios, e.g., answering questions, writing code, etc. 
{\tool} can also generate some text with actual meaning, such as \textit{``Blood pressure record''} in Fig \ref{fig:RQ1-goodcase} (a).
% , i.e., \textit{``My name is Franklin, my personality is bright, honest, easy to get along with people. I like playing basketball.''}. \chen{Not a very good example here, more appropriate to put into RQ2. I suggest using the example of flight/product search, which are filling in blanks.}
% Since {\tool} is based on a large language model, as shown in the Fig \ref{fig:RQ1-goodcase} (b) \& (c), it can cover more scenarios, such as answering questions, writing code, etc. {\tool} can also generate some text with actual meaning, such as ``describe yourself'' in Fig \ref{fig:RQ1-goodcase} (a).
We further examine the cases for the un-passing page of {\tool} and summarize the following two reasons. (1) GUI pages do not have contextual widgets and text input without semantic information. 
Incorporating other contextual information as the nearby UI pages can enhance the understanding of the query widget in such cases. 
% It may be needed to analyze the information on nearby UI pages.
(2) Apps need specific and unique text input, such as server address, database connection, etc.

We also investigate the contribution of prompt-based data construction and tuning method by comparing the passing rate on the overall with prompt-tuning data.
% and \chen{Is this sentence complete? Is this the ablation study?}
% \rev{on the {\tool}} without prompt-tuning (details are in Section \ref{subsec_experiment_dataset}).
% \jie{comment}
%IV-A现在没有这部分内容了。
From Table \ref{tab:RQ1-result}, we can see that passing rate improve by 43\% (0.61 vs 0.87) when the tuning data is used to prompt-tune the model, indicating the value of automated generated tuning data for effective the input text generation. 
% \rev{Specifically, 43\% (0.61 vs 0.87) improvement is observed respectively for passing rate. }
% \rev{With the extracted tuning dataset, input text that has a higher passing rate can be generated.}

\subsubsection{\textbf{Input Text Generation Quality} (RQ2)}
\label{sec_results_RQ2}
\begin{figure}[htb]
\centering
\vspace{-0.05in}
\includegraphics[width=8.3cm]{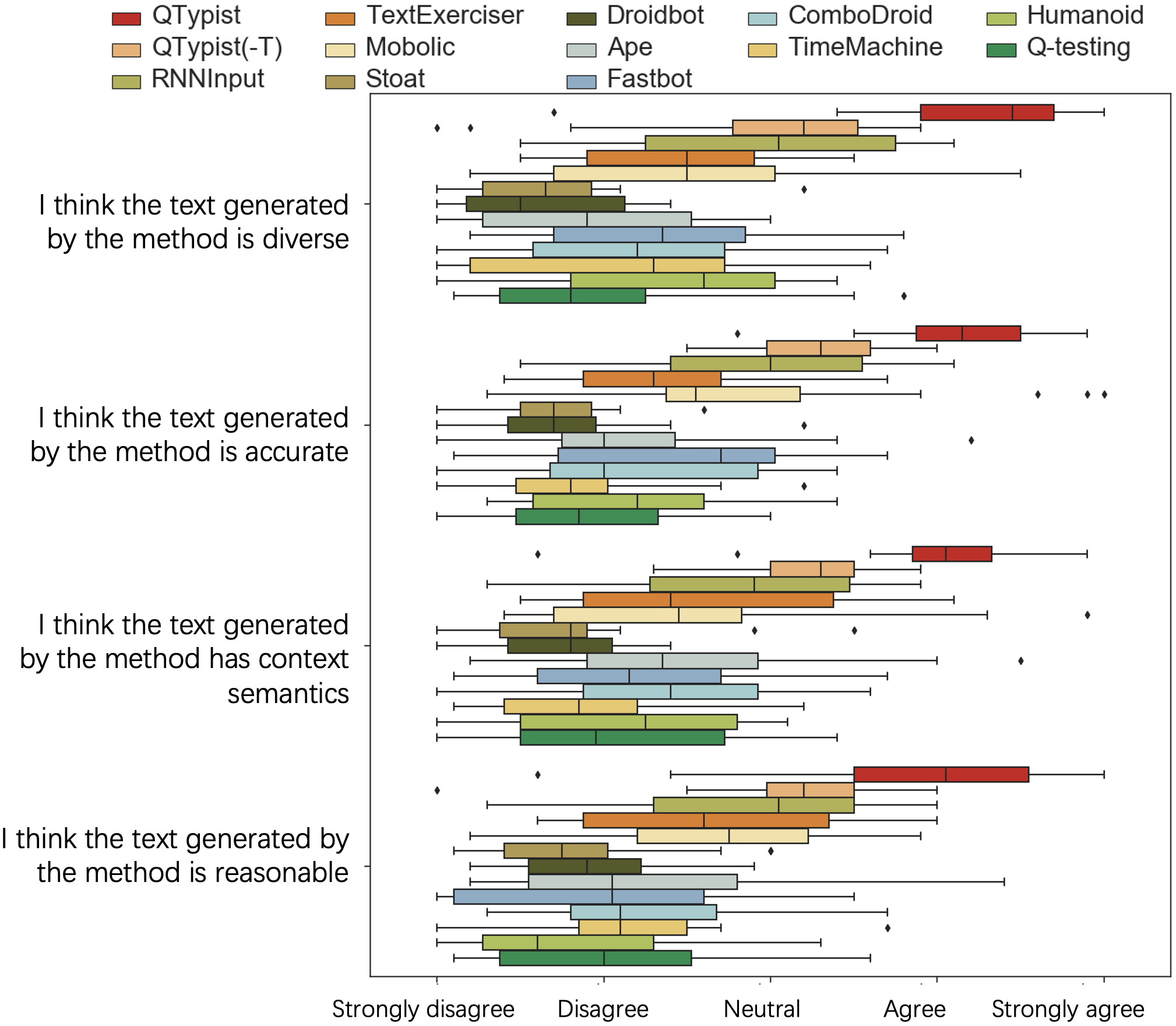}
% \vspace{-0.1in}
\caption{The result of the accuracy of generated input.}
\label{fig:RQ2-5-likert}
\vspace{-0.1in}
\end{figure}

Fig \ref{fig:RQ2-5-likert} provides the responses to the 5-Likert scale from the 20 testers. Overall, the responses from these testers are quite positive. The practitioners strongly agree or agree with the diversity and the accuracy of input content generation results by {\tool}, i.e., the average score is 4.4. They generally agree the input content generated with {\tool} is easy to understand, and more reasonable. 
The result of average Kendall's W~\cite{Kendall}
% \chen{What is this? If not mentioned above, please explain}
% \liuzhe{comment}
% 我这边在实验设置方面RQ2部分介绍了一下
is 0.8, which indicates a high degree of inter-agreement on the performance of {\tool}'s input content generation.

% Additionally, we highlight representative responses \jie{where is the response?} to the 5-Likert scale in this section, 
% more detailed information can be found in our website\textsuperscript{\ref{github}}.

% \jie{comment}
%Qtypist(without) 和上面章节保持一致吧
%感觉figure 6给出的例子都是短的，对于那种comments相关的，要给出个例子吗，这里可以是其他方法也能通过的，但是我们的生成比较有semantic，正好可以在这里提一下。
%但是我觉得需要给出我们生成有意义的输入的优势，如果都是能通过，只是我们的有意义，好像也挺弱的，有没有可能后续有利于帮助理解，或者后续有啥潜在作用或者啥的。我觉得这个层面的如果能有个例子更好了

\begin{table*}
\vspace{-0.13in}
\renewcommand\arraystretch{0.95} 
\caption{Result of activity and page number compare with automated GUI testing tool with {\tool}.} 
% \chen{for the last row, please also add the average number.}}
\vspace{-0.05in}
\label{tab:Result-usefulness}
\centering
\footnotesize
\begin{tabular}{p{0.3cm}<{\centering} | p{1.2cm}<{\centering} | p{0.8cm}<{\centering} | p{0.7cm}<{\centering} | p{0.2cm}<{\centering} | p{1.05cm}<{\centering} || p{0.25cm}<{\centering} | p{1.05cm}<{\centering} || p{0.25cm}<{\centering} | p{1.05cm}<{\centering} || p{0.25cm}<{\centering} | p{1.05cm}<{\centering} || p{0.25cm}<{\centering} | p{1.05cm}<{\centering} || p{0.25cm}<{\centering} | p{1.05cm}<{\centering}}
% \hline
\toprule
\multirow{2}*{\textbf{Id}} & \multirow{2}*{\textbf{App name}} & \multirow{2}*{\textbf{Categ}}  & \multirow{2}*{\textbf{Down}} & \multicolumn{6}{c||}{\textbf{Triggered activity number}} & \multicolumn{6}{c}{\textbf{Triggered page number}} \cr 
% \cline{5-16}
 &  &  &  & \textbf{M} & \textbf{M+QT} & \textbf{D} & \textbf{D+QT} & \textbf{A} & \textbf{A+QT} & \textbf{M} & \textbf{M+QT} & \textbf{D} & \textbf{D+QT} & \textbf{A} & \textbf{A+QT}\\
% \hline
\midrule[0.5pt]
1 & Eski & Social & 1M+ & 3 & \ 4 \textcolor{red}{$\uparrow$}33\%  & \ 4 & \ 5 \textcolor{red}{$\uparrow$}25\%  & \ 4 & \ 6 \textcolor{red}{$\uparrow$}50\%  & \ 5 & \ 8 \textcolor{red}{$\uparrow$}60\%  & \ 7 & 10 \textcolor{red}{$\uparrow$}43\%  & \ 8 & 14 \textcolor{red}{$\uparrow$}75\% \\ \rowcolor{gray!15}
2 & Recha & Shop & 1M+ & 5 & \ 5 \textcolor{green}{ - }\ 0\%  & \ 8 & 13 \textcolor{red}{$\uparrow$}63\%  & \ 9 & 15 \textcolor{red}{$\uparrow$}67\%  & \ 6 & \ 8 \textcolor{red}{$\uparrow$}33\%  & 10 & 16 \textcolor{red}{$\uparrow$}60\%  & 11 & 18 \textcolor{red}{$\uparrow$}64\% \\ 
3 & Scout & Vehicle & 10M+ & 4 & \ 7 \textcolor{red}{$\uparrow$}75\%  & 10 & 17 \textcolor{red}{$\uparrow$}70\%  & 12 & 19 \textcolor{red}{$\uparrow$}58\%  & \ 5 & \ 9 \textcolor{red}{$\uparrow$}80\%  & 11 & 19 \textcolor{red}{$\uparrow$}73\%  & 14 & 24 \textcolor{red}{$\uparrow$}71\% \\ \rowcolor{gray!15}
4 & Stoc & Finance & 1M+ & 5 & \ 9 \textcolor{red}{$\uparrow$}80\%  & 12 & 18 \textcolor{red}{$\uparrow$}50\%  & 10 & 18 \textcolor{red}{$\uparrow$}80\%  & \ 7 & 13 \textcolor{red}{$\uparrow$}86\%  & 15 & 23 \textcolor{red}{$\uparrow$}53\%  & 16 & 26 \textcolor{red}{$\uparrow$}63\% \\ 
5 & Lit & Photo & 1M+ & 4 & \ 6 \textcolor{red}{$\uparrow$}50\%  & \ 6 & 11 \textcolor{red}{$\uparrow$}83\%  & \ 9 & 12 \textcolor{red}{$\uparrow$}33\%  & \ 5 & \ 8 \textcolor{red}{$\uparrow$}60\%  & 12 & 19 \textcolor{red}{$\uparrow$}58\%  & 14 & 23 \textcolor{red}{$\uparrow$}64\% \\ \rowcolor{gray!15}
6 & Speedo & Vehicle & 1M+ & 4 & \ 5 \textcolor{red}{$\uparrow$}25\%  & \ 8 & \ 8  \textcolor{green}{ - }\ 0\%  & \ 8 & 13 \textcolor{red}{$\uparrow$}63\%  & \ 5 & \ 6 \textcolor{red}{$\uparrow$}20\%  & \ 9 & \ 9 \textcolor{green}{ - }\ 0\%  & \ 9 & 15 \textcolor{red}{$\uparrow$}67\% \\ 
7 & HeartR & Health & 1M+ & 5 & \ 5 \textcolor{green}{ - }\ 0\%  & \ 6 & \ 7 \textcolor{red}{$\uparrow$}17\%  & \ 6 & \ 9 \textcolor{red}{$\uparrow$}50\%  & \ 6 & \ 6  \textcolor{green}{ - }\ 0\%  & \ 7 & \ 9 \textcolor{red}{$\uparrow$}29\%  & \ 8 & 14 \textcolor{red}{$\uparrow$}75\% \\ \rowcolor{gray!15}
8 & Curren & Tools & 1M+ & 3 & \ 3 \textcolor{green}{ - }\ 0\%  & \ 5 & \ 7 \textcolor{red}{$\uparrow$}40\%  & \ 6 & \ 7 \textcolor{red}{$\uparrow$}17\%  & \ 4 & \ 4  \textcolor{green}{ - }\ 0\%  & \ 6 & \ 8 \textcolor{red}{$\uparrow$}33\%  & \ 7 & 11 \textcolor{red}{$\uparrow$}57\% \\ 
9 & Yande & Shop & 10M+ & 8 & 11 \textcolor{red}{$\uparrow$}38\%  & 14 & 20 \textcolor{red}{$\uparrow$}43\%  & 15 & 22 \textcolor{red}{$\uparrow$}47\%  & 13 & 16 \textcolor{red}{$\uparrow$}23\%  & 17 & 23 \textcolor{red}{$\uparrow$}35\%  & 21 & 31 \textcolor{red}{$\uparrow$}48\% \\ \rowcolor{gray!15}
10 & Atom & Finance & 1M+ & 2 & \ 2 \textcolor{green}{ - }\ 0\%  & \ 3 & \ 4 \textcolor{red}{$\uparrow$}33\%  & 3 & \ 5 \textcolor{red}{$\uparrow$}67\%  & \ 4 & \ 4  \textcolor{green}{ - }\ 0\%  & \ 5 & \ 7 \textcolor{red}{$\uparrow$}40\%  & \ 5 & \ 8 \textcolor{red}{$\uparrow$}60\% \\ 
11 & Coco & Comm & 10M+ & 2 & \ 3 \textcolor{red}{$\uparrow$}50\%  & \ 8 & 13 \textcolor{red}{$\uparrow$}63\%  & \ 7 & 13 \textcolor{red}{$\uparrow$}86\%  & 3 & \ 5 \textcolor{red}{$\uparrow$}67\%  & 10 & 15 \textcolor{red}{$\uparrow$}50\%  & \ 9 & 15 \textcolor{red}{$\uparrow$}67\% \\ \rowcolor{gray!15}
12 & Savir & Shop & 100K+ & 4 & \ 7 \textcolor{red}{$\uparrow$}75\%  & \ 8 & 12 \textcolor{red}{$\uparrow$}50\%  & \ 9 & 13 \textcolor{red}{$\uparrow$}44\%  & \ 8 & 12 \textcolor{red}{$\uparrow$}50\%  & 15 & 23 \textcolor{red}{$\uparrow$}53\%  & 18 & 29 \textcolor{red}{$\uparrow$}61\% \\ 
13 & WhiteM & Dating & 100K+ & 7 & \ 8 \textcolor{red}{$\uparrow$}14\%  & 10 & 13 \textcolor{red}{$\uparrow$}30\%  & 10 & 16 \textcolor{red}{$\uparrow$}60\%  & 10 & 12 \textcolor{red}{$\uparrow$}20\%  & 15 & 19 \textcolor{red}{$\uparrow$}27\%  & 16 & 25 \textcolor{red}{$\uparrow$}56\% \\ \rowcolor{gray!15}
14 & Healthp & Health & 1M+ & 5 & \ 5  \textcolor{green}{ - }\ 0\%  & \ 7 & \ 7  \textcolor{green}{ - }\  0\%  & \ 8 & 13 \textcolor{red}{$\uparrow$}63\%  & \ 9 & 10 \textcolor{red}{$\uparrow$}11\%  & 11 & 13 \textcolor{red}{$\uparrow$}18\%  & 13 & 19 \textcolor{red}{$\uparrow$}46\% \\ 
15 & LINE & Photo & 10M+ & 5 & \ 7 \textcolor{red}{ $\uparrow$}40\%  & 12 & 17 \textcolor{red}{$\uparrow$}42\%  & 11 & 18 \textcolor{red}{$\uparrow$}64\%  & \ 6 & \ 9 \textcolor{red}{$\uparrow$}50\%  & 11 & 19 \textcolor{red}{$\uparrow$}73\%  & 15 & 27 \textcolor{red}{$\uparrow$}80\% \\ \rowcolor{gray!15}
16& FlPla & Art& 5M+& 4& \ 5 \textcolor{red}{$\uparrow$}25\%& \ 7& \ 7 \textcolor{green}{ - }\ 0\%& \ 9& 10 \textcolor{red}{$\uparrow$}11\%& \ 5& \ 8 \textcolor{red}{$\uparrow$}60\%& 10& 11 \textcolor{red}{$\uparrow$}10\%& 13& 18 \textcolor{red}{$\uparrow$}38\%\\ 
17& MoneyTK& Finance& 10M+& 4& \ 6 \textcolor{red}{$\uparrow$}50\%& \ 8& \  9 \textcolor{red}{$\uparrow$}13\%& \ 9& 13 \textcolor{red}{$\uparrow$}44\%& \ 7& \ 8 \textcolor{red}{$\uparrow$}14\%& 15& 20 \textcolor{red}{$\uparrow$}33\%& 17& 26 \textcolor{red}{$\uparrow$}53\%\\ \rowcolor{gray!15}
18& Schoolca & Educat& 1M+& 2& \ 2 \textcolor{green}{ - }\ 0\%& \ 4& \ 6 \textcolor{red}{$\uparrow$}50\%& \ 8& 17 \textcolor{red}{$\uparrow$}50\%& \ 4& \ 4 \textcolor{green}{ - }\ 0\%& \ 8& 10 \textcolor{red}{$\uparrow$}25\%& \ 8& 13 \textcolor{red}{$\uparrow$}63\%\\ 
19& Flipbo & News& 50M+& 3& \ 4 \textcolor{red}{$\uparrow$}33\%& \ 4& \ 5 \textcolor{red}{$\uparrow$}25\%& \ 7& \ 7 \textcolor{green}{ - }\ 0\%& \ 5& \ 8 \textcolor{red}{$\uparrow$}60\%& \ 7& 11 \textcolor{red}{$\uparrow$}57\%& \ 8& 12 \textcolor{red}{$\uparrow$}50\%\\ \rowcolor{gray!15}
20& Healtp & Fitness& 1M+& 4& \ 4 \textcolor{green}{ - }\ 0\%& \ 8& \ 9 \textcolor{red}{$\uparrow$}13\%& 11& 12 \textcolor{red}{$\uparrow$ }9\%& \ 6& \ 6 \textcolor{green}{ - }\ 0\%& 13& 19 \textcolor{red}{$\uparrow$}46\%& 15& 20 \textcolor{red}{$\uparrow$}33\%\\ 
21& BlaW & Travel& 1M+& 5& \ 6 \textcolor{red}{$\uparrow$}20\%& 12& 14 \textcolor{red}{$\uparrow$}17\%& 15& 17 \textcolor{red}{$\uparrow$}13\%& \ 5& \ 6 \textcolor{red}{$\uparrow$}20\%& 18& 25 \textcolor{red}{$\uparrow$}39\%& 21& 27 \textcolor{red}{$\uparrow$}29\%\\ \rowcolor{gray!15}
22& Wldetect & Product& 1M+& 3& \ 3 \textcolor{green}{ - }\ 0\%& \ 8& \ 9 \textcolor{red}{$\uparrow$}13\%& 12& 16 \textcolor{red}{$\uparrow$}33\%& \ 5& \ 6 \textcolor{red}{$\uparrow$}20\%& 15& 16 \textcolor{red}{$\uparrow$ }7\%& 17& 18 \textcolor{red}{$\uparrow$ }6\%\\ 
23& CrAm & Educat& 5M+& 2& \ 3 \textcolor{red}{$\uparrow$}50\%& \ 4& \ 7 \textcolor{red}{$\uparrow$}75\%& \ 9& 10 \textcolor{red}{$\uparrow$}11\%& \ 7& \ 7 \textcolor{green}{ - }\ 0\%& \ 8& 15 \textcolor{red}{$\uparrow$}88\%& 11& 16 \textcolor{red}{$\uparrow$}45\%\\ \rowcolor{gray!15}
24& InsTE & Game& 100K+& 4& \ 5 \textcolor{red}{$\uparrow$}25\%& \ 7& \ 8 \textcolor{red}{$\uparrow$}14\%& \ 9& 12 \textcolor{red}{$\uparrow$}33\%& \ 6& \ 9 \textcolor{red}{$\uparrow$}50\%& 11& 19 \textcolor{red}{$\uparrow$}73\%& 10& 19 \textcolor{red}{$\uparrow$}90\%\\ 
25& FitN & Connect& 1M+& 7& \ 8 \textcolor{red}{$\uparrow$}14\%& 12& 15 \textcolor{red}{$\uparrow$}25\%& 15& 18 \textcolor{red}{$\uparrow$}20\%& \ 8& 10 \textcolor{red}{$\uparrow$}25\%& 17& 25 \textcolor{red}{$\uparrow$}47\%& 19& 29 \textcolor{red}{$\uparrow$}53\%\\ \rowcolor{gray!15}
26& GPST & Navig& 1M+& 4& \ 4 \textcolor{green}{ - }\ 0\%& \ 6& \ 8 \textcolor{red}{$\uparrow$}33\%& \ 8& 11 \textcolor{red}{$\uparrow$}38\%& \ 6& \ 6 \textcolor{green}{ - }\ 0\%& 10& 14 \textcolor{red}{$\uparrow$}40\%& \ 9& 13 \textcolor{red}{$\uparrow$}44\%\\ 
27& DMCR & News& 100K+& 5& \ 7 \textcolor{red}{$\uparrow$}40\%& \ 6& \ 8 \textcolor{red}{$\uparrow$}33\%& \ 8& 10 \textcolor{red}{$\uparrow$}25\%& 10& 13 \textcolor{red}{$\uparrow$}30\%& 11& 14 \textcolor{red}{$\uparrow$}27\%& 10& 14 \textcolor{red}{$\uparrow$}40\%\\ \rowcolor{gray!15}
28& Met & Navig& 1M+& 3& \ 3 \textcolor{green}{ - }\ 0\%& \ 4& \ 4 \textcolor{green}{ - }\ 0\%& \ 6& \ 8 \textcolor{red}{$\uparrow$}33\%& \ 5& \ 5 \textcolor{green}{ - }\ 0\%& \ 6& \ 6 \textcolor{green}{ - }\ 0\%& \ 7& \ 9 \textcolor{red}{$\uparrow$}29\%\\ 
29& Wall & Finance& 5M+& 4& \ 6 \textcolor{red}{$\uparrow$}50\%& \ 5& \ 8 \textcolor{red}{$\uparrow$}60\%& \ 7& 10 \textcolor{red}{$\uparrow$}43\%& \ 4& \ 6 \textcolor{red}{$\uparrow$}50\%& \ 6& \ 9 \textcolor{red}{$\uparrow$}50\%& \ 9& 12 \textcolor{red}{$\uparrow$}33\%\\ \rowcolor{gray!15}
30& Pock & Travel& 100K+& 4& \ 5 \textcolor{red}{$\uparrow$}25\%& \ 6& \ 7 \textcolor{red}{$\uparrow$}17\%& \ 9& 12 \textcolor{red}{$\uparrow$}33\%& \ 7& \ 8 \textcolor{red}{$\uparrow$}14\%& \ 9& 11 \textcolor{red}{$\uparrow$}22\%& 12& 16 \textcolor{red}{$\uparrow$}33\%\\ 

% \hline
\midrule
\multicolumn{4}{c|}{\textbf{Average boost
}} & \multicolumn{2}{c||}{\textcolor{red}{$\uparrow$}\ \textbf{28\%}} & \multicolumn{2}{c||}{\textcolor{red}{$\uparrow$}\ \textbf{33\%}} & \multicolumn{2}{c||}{\textcolor{red}{$\uparrow$}\ \textbf{42\%}} & \multicolumn{2}{c||}{\textcolor{red}{$\uparrow$}\ \textbf{30\%}} & \multicolumn{2}{c||}{\textcolor{red}{$\uparrow$}\ \textbf{41\%}} & \multicolumn{2}{c}{\textcolor{red}{$\uparrow$}\ \textbf{52\%}} \\
% \multicolumn{4}{c|}{\textbf{Total}} & & \textcolor{red}{$\uparrow$}\ \textbf{28\%} & &  \textcolor{red}{$\uparrow$}\ \textbf{33\%} & & \textcolor{red}{$\uparrow$}\ \textbf{42\%} & &  \textcolor{red}{$\uparrow$}\ \textbf{30\% }& &  \textcolor{red}{$\uparrow$}\ \textbf{41\%} & &  \textcolor{red}{$\uparrow$}\ \textbf{52\%} \\
% & & & & 124& 159 \textcolor{red}{$\uparrow$}28\%& 222& 296 \textcolor{red}{$\uparrow$}33\%& 269& 382 \textcolor{red}{$\uparrow$}42\%& 186& 241 \textcolor{red}{$\uparrow$}30\%& \ 325& 457 \textcolor{red}{$\uparrow$}41\%& \ 370& 561 \textcolor{red}{$\uparrow$}52\%\\
% \hline
\bottomrule
\end{tabular}
\vspace{0.05in}
\begin{tablenotes}
\footnotesize
\item \textbf{\textit{Notes:}} ``M'' is Monkey, ``D'' is DroidBot, ``A'' is Ape. ``M+QT'', ``D+QT'', ``A+QT'' are Monkey, DroidBot and Ape with {\tool}. ``Categ'' is app categories. ``Down'' is download number. ``\textcolor{red}{$\uparrow$}'' means performance increase of automated testing tools after integrating {\tool} and ``\textcolor{green}{-}'' means no growth.
\end{tablenotes}
% \vspace{ \textcolor{green}{ - }\ 0.05in}
\vspace{-0.15in}
\end{table*}

We further analyze the responses of the baselines and {\tool} without fine-tuning. As shown in Fig \ref{fig:RQ2-5-likert}, participants have a certain degree of recognition for the input text generated by {\tool}, the average score is 4.2. 
% \jie{comment}
%这里只看这句话，觉得没有fine tune好像也还挺好的，你应该说这个比整个方法要差不好。
%后面那句，have certain meaning也感觉好像还行的样子，看看稍微变个说法
For the other 11 baseline methods, participants believe that the text generated by RNNInput and TextExerciser 
% only have certain meaning and was 
is incorrect in many cases. However, participants disagreed or strongly disagreed with the text generated by other automated GUI testing tools. Participants believe that the text they randomly generated has no semantics and practical meaning even if it could be passed by chance. As shown in the upper part of Fig \ref{fig:RQ1-goodcase}, although the RNNInput and TextExerciser can also pass the test, the input text does not conform to the actual semantics. {\tool} generates text with semantics, which helps to find more potential issues, e.g., display issues or compatibility issues caused by long input text.

\section{Usefulness Evaluation}
\label{sec_Usefulness}
In addition to the effectiveness evaluation in the previous section, we further evaluate the usefulness of {\tool} in augmenting the automated GUI testing. 
Our goal is to examine: (1) Whether {\tool} can effectively help the automated GUI testing tools to cover more UI pages? 
% \jie{comment}
%这里是不是直接说覆盖更多页面就可以？
(2) Whether {\tool} can facilitate the automated GUI testing tools to find more bugs (since more pages are explored)?
% (3) whether {\tool} can save the testing time?

\subsubsection{\textbf{Experiment Setup}}
\label{sub_Usefulness_automated_Experiment}
% We modify the automated GUI testing tool to enable it potentially explore more states and find more bugs.
% \jie{comment}
%逻辑不对吧，modify it 是为了整合上我们的工具，至于是不是more states 也不是 在这里说的逻辑
We randomly sample 30 Android apps from Google play, which are not covered in effectiveness evaluation in Section \ref{sec_Effectiveness}, and the 3 automated GUI testing tools can run these apps normally.
They have at least one text input, with source code and issue reports on GitHub.
For some apps that require actual account login, we skip the process through scripts.
We use the number of triggered activities and UI pages~\cite{su2017guided,he2020textexerciser}, and the number of revealed bugs as our metrics.
Note that there may be multiple UI pages in an activity, so we also use the triggered UI pages for evaluation. 
The judgment basis of UI pages accords with previous studies~\cite{su2017guided,yang2018static}.
We integrate {\tool} into 3 popular and commonly-used automated GUI testing tools, Monkey, DroidBot and Ape, as examples for usefulness evaluation.
All experiments are conducted on the official Android x64 emulator running Android 6.0 on a server with Xeon E5-2650 v4 processors. Each emulator is allocated with 4 dedicated CPU cores, 4 GB of RAM, and 4 GB of internal storage space.

\subsubsection{\textbf{Result}}
\label{subsub_Effectiveness_Result}
Table \ref{tab:Result-usefulness} shows the results.
On average, the triggered activity and page number of the automated GUI testing tools with {\tool} are higher than those without {\tool}. Particularly, Monkey with {\tool} triggers 28\% more activities and 30\% more pages compared than Monkey, DroidBot with {\tool} triggers 33\% more activities and 41\% more pages than DroidBot, Ape with {\tool} triggers 42\% more activities and 52\% more pages than Ape.
% \jie{comment}
%这些数字不太对吧，好好对一下。

% Figure \ref{fig:usefulness-a-b} shows the activity coverage of each app which is tested by the three with the automated GUI testing tools integrated our {\tool} and they without our {\tool}, under different testing time settings (10, 20, 30 minutes).
% The experimental results show that the three automated testing tools integrated with our method have achieved high activity coverage and page coverage. Among them, Ape with {\tool} shows a good effect.

% \input{figure/usefulness-a-b}

% For activity coverage, Ape with integrated our {\tool} achieves a median activity coverage from each app of 63\% with the range from 0.37 to 1 across 30 mobile apps (each app with 30 minutes running time), which is 53\% (63\% vs 41\%) more than the activiy coverage of Ape without our {\tool}. 
% This indicates the effectiveness of the input content generated by our {\tool} in augmenting automated GUI testing tools, i.e., generating the correct input text to realize the successful transition between the states. 

% \jie{comment}
%这个数据只有Ape，没有其他的了？我觉得可以不用列平均，就直接列出总数吧，每个工具，用或者没用Qtypist的总数，以及增长比例。

% \rev{Through armed with {\tool}, these automated GUI testing tools can cover more UI pages and states, which could potentially reveal more bugs. 
% We then present the number of detected bugs with {\tool} compared the raw tool. Note that, these bugs are the ones already recorded in the app's issue repositories, and in next paragraph, we will illustrate the capability of {\tool} in help finding new bugs.
% In detail, }
With {\tool}, the Ape can find 51 bugs from the 30 apps, which is 122\% (51 vs 23) more than the bugs detected by Ape without {\tool}. These bugs are crash bugs that have been confirmed by the developers.
% \rev{ but not fixed yet. }
Monkey with {\tool} detects 109\% (23 vs 11) more bugs and DroidBot with {\tool} detects 106\% (35 vs 17) more bugs than them without {\tool}.
% \rev{Through further analysis, we find that Ape cannot detect the bugs with input or require input operations to trigger.}
It further indicates the practical value of {\tool}.

\section{Discussion and Threats to Validity}
\label{sec_discussion}

\subsection{Discussion}
% \chen{Remove it if more space is needed.}
\subsubsection{\textbf{Generality Across Platforms}}
Almost all existing studies of input text generation tools and automated GUI testing tools~\cite{zein2016systematic, lamsa2017comparison, DroidBot, he2020textexerciser,liu2017automatic} are designed for a specific platform, e.g., Android, which limits its applicability in real-world practice.
In comparison, the primary idea of {\tool} is to generate input content from input widgets and contextual information when running the apps.
Since the different platforms have these similar types of information, {\tool} can be used to generate input content in other platforms.
We have conducted a small-scale experiment for another two popular platforms, i.e., iOS and Web, and experimented on 20 apps of each platform.
Results show that {\tool} can pass the UI page in 90\% apps from iOS and 85\% apps from Web.
This further demonstrates the generality of {\tool}, and we will conduct more thorough experiments in the future.

\subsubsection{\textbf{Generality Across Languages}}
Another advantage of {\tool} is that it can generate the input content in terms of different display languages of the app.
We also collect 50 apps in 5 other languages (i.e. German, French, Italian, Chinese and Korean) from Google Play, and run {\tool} to generate input content.
Note that {\tool} is prompt-tuned with English language apps as described in Section \ref{subsec_approach_Implement}. 
Results show that {\tool} can accurately pass UI pages requiring text inputs in 86\% apps,
% the UI page in 86\% apps,
% which further demonstrates its feasibility.
which further demonstrates the feasibility of {\tool}.

\subsubsection{\textbf{Assisting Manual Testing}}
We demonstrate {\tool} in helping with the automated GUI testing in previous sections.
However, {\tool} is not limited to automated GUI testing, and it can also be utilized for assisting manual testing, e.g., suggests diversified test cases of input widget for crowdtesters. 
% \rev{It can provide more abundant test cases of text inputs for the crowdtesters.}
As the complexity of apps increases, crowdtesters often do not know what to input when testing unfamiliar apps. 
If {\tool} can provide diversified test cases for the crowdtesters, it can potentially help the crowdtesters find more bugs.

\subsection{Threats of Validity}
\label{subsec_Validity}
% The external threat to the validity of our work is the  environmental dependencies of our experimental apps. Some of the experimental apps in our experiments require networking for main functionalities to be usable, and it is possible for such dependency to change the behaviors of these apps despite our efforts to make our experiment environment consistent across different runs.

The external threat to the validity of our work is the representativeness and generality of our experimental apps in Section \ref{sec_Effectiveness} and \ref{sec_Usefulness}. To reduce this threat, we interview 4 industrial practitioners with more than 8 years of testing experience to obtain the agreed-upon selection criterion of apps. Table \ref{tab:app-info} and \ref{tab:Result-usefulness} shows the apps are diverse.

% Some of the experimental apps in our experiments require networking for main functionalities to be usable, and it is possible for such dependency to change the behaviors of these apps despite our efforts to make our experiment environment consistent across different runs.

% The internal threat to the validity of our work comes from the manual analysis of the input content generated from our {\tool} and other baselines. We need to manually determine whether the content generated by {\tool} and other baselines satisfies the actual requirements and has semantic information. Consequently, related evaluation results can be influenced by subjective judgments. However, it should be noted that any work involving manual judgments in the evaluation is vulnerable to this threat. To counter this, we make considerable effort to set a rigorous experimental setup, and invite 20 experienced testers to evaluate. (see Section \ref{sec_results_RQ2}).

The internal threat to the validity of our work comes from the process of manual inspection and tagging. Related evaluation results can be influenced by subjective judgments. We understand that such a process is subject to mistakes, and it should be noted that any work involving manual judgments in evaluation is vulnerable to this threat. To reduce it, we build an inspection team to reach agreements on different options.

% manual analysis of the input content generated from our {\tool} and other baselines. We need to manually determine whether the content generated by {\tool} and other baselines satisfies the actual requirements and has semantic information. Consequently, related evaluation results can be influenced by subjective judgments. However, it should be noted that any work involving manual judgments in the evaluation is vulnerable to this threat. To counter this, we make considerable effort to set a rigorous experimental setup, and invite 20 experienced testers to evaluate. (see Section \ref{sec_results_RQ2}).

\section{Related Work}
\label{sec_related}
\subsection{Automated GUI testing}
To ensure the quality of mobile apps, many researchers study the automatic generation of large-scale test scripts to test apps~\cite{xie2007designing}.
Since Android apps are event-based~\cite{anand2012automated,wu2019analyses,jabbarvand2019search,matinnejad2017automated}, the most common automated testing methods are model-based~\cite{mirzaei2016reducing,yang2018static,yang2013grey}.
Some linting tools~\cite{lint,stylelint} based on static program analysis to mark programming errors, bugs.
However, due to the complexity of the mobile apps, random GUI testing tools based on the dynamic analysis are proposed~\cite{machiry2013dynodroid,zeng2016automated,mao2016sapienz} with automatic exploration, which aim at covering more pages or activities. 
These dynamic automated GUI testing tools~\cite{borges2018droidmate,li2019humanoid, su2017guided,Monkey,li2017droidbot,white2019improving,degott2019learning,kowalczyk2018configurations,moran2018mdroid+,cruz2019energy} simulated human operation (click, slide, etc.) to test the app~\cite{wang2018empirical}.
Although they can test apps by designing exploration strategies with random actions, they can't handle the complex operations like text input with context semantics, resulting in low activity coverage. Some researches~\cite{zein2016systematic,lamsa2017comparison,guo2020improving,wang2021infrastructure} tried to improve the coverage of automated testing tools by improving the exploration algorithm.
This study focuses on a different direction for improving the activity coverage, i.e., generating suitable input text.
% \rev{, they still performed poorly on some apps with some text input.
% Therefore, we propose an approach to help automated GUI testing tools generate input text. }

\subsection{Text Input Generation}
% \rev{In Android apps, researchers pay less attention to automatic text input generation. }
The automated text input generation of mobile apps are not well explored by existing studies. 
Monkey~\cite{Monkey} and Dynodroid~\cite{machiry2013dynodroid}, only can generate UI events like randomly clicking the button but they can not generate input content.
% \jie{comment}
%input context 是啥？
Sapienz~\cite{mao2016sapienz},
Stoat~\cite{su2017guided},
A3E-Depth-First~\cite{azim2013targeted}, DroidBot~\cite{DroidBot}, AppsPlayground~\cite{rastogi2013appsplayground}, Ape~\cite{gu2019practical}, Fastbot~\cite{cai2020fastbot}, ComboDroid~\cite{wang2020combodroid}, TimeMachine~\cite{dong2020time}, Humanoid~\cite{li2019humanoid} and DroidDEV~\cite{arnatovich2016achieving}, fulfilled the text input by searching in a set of pre-defined candidates. 
If none of the pre-defined candidates  can satisfy an input’s constraints, these prior works will fail to exercise beyond this input. Liu et al. utilized RNNInput~\cite{liu2017automatic} to generate text input based on the app input information. 
% \jie{comment}
%based on the app input information 是啥。related work这块我没怎么细看，这块我不看了，质量你自己把握哈
Unfortunately, it required a large amount of manual effort to write input for training and didn't consider context. TextExerciser~\cite{he2020textexerciser} identified the input restrictions from UI screen by UI structural analysis and then generated a text input with a mutation based strategy. 
In Web apps, SWAT~\cite{alshahwan2011automated}, AWET~\cite{sunman2022automated} generated the input test based on pre-defined template.
ACTEve~\cite{anand2012automated} and S3~\cite{trinh2014s3} first use symbolic execution to extract input constraints in the source code and then use a solver to generate an input. They need to analyze the web code and can't be directly applied to Android apps with complex rendering mechanisms.
Furthermore, {\tool} can potentially be utilized for the input generation of web apps.

\subsection{Large Language Model and Prompt Engineering}
Recently, the great success of pre-trained Large Language Models (e.g., BERT~\cite{devlin2018bert,sanh2019distilbert}, RoBERTa~\cite{2019RoBERTa}, GPT-3~\cite{brown2020GPT3}, T5~\cite{DBLP:journals/jmlr/RaffelSRLNMZLL20}) in a variety of NLP tasks. 
% \jie{comment}
%这里我们需要重新引入PLM这个缩写吗
Researchers study how to improve the robustness and performance of models in different fields through prompt engineering~\cite{DBLP:journals/corr/abs-2207-11680,liao2022ptau,chen2022knowprompt,zhou2022learning,wu2022fast,wang2022promda,cui2022prototypical,gu2021ppt,liu2022p}. 
Gwern branwen et al.~\cite{Branwen2020Gpt-3creative, Cantino201Prompt} proposed that the prompt engineering model can become a new interaction paradigm. Users only needed to know how to prompt the model to obtain the specific knowledge and abstractions needed to complete downstream tasks.
Liu et al.~\cite{liu2021pre} formalized a paradigm, which classifies prompts according to the shape of prompts (cloze prompts), answer engineering (answered prompts) and task specific prompts. 
% They also proposed multi prompt engineering for prompt tuning.
% In addition, they also extended the alternative methods of prompt engineering, such as automatic template learning and multi prompt engineering.
Ge et al.~\cite{ge2021visual} used BigSleep and DeepDaze to integrate visual concepts. They used BERT to generate prompts, helped users with visual fusion, and used shape keywords to initialize the generation.
Aran komatsuzaki~\cite{Komatsuzaki2021When} pointed out that using the ``unreal engine'' as a hint helped them add surreal 3D rendering quality to image generation.
% So far, the research on prompt
%这里需要引入这个新的名词吗。感觉这句话可以去掉吧，已经有些progress也没啥需要说的。就直接我们做了什么就行。
%我们是生成summary information 吗？
% engineering for different scenarios. 
This paper uses context-aware input prompt generation method to generate the prompt of input widgets.

\section{Conclusion}
\label{sec_conclusion}
Automated GUI testing is crucial for helping improve app quality.
However, text input generation remains challenging, which hinders the large-scale adoption of automated testing approaches. 
We propose {\tool} which adopts pre-trained Large Language Model to automatically generate input text. 
It automatically generates the text input prompt related to the text inputs' contextual information as the LLM input. 
To boost the performance, we develop a prompt-based data construction and tuning method which automatically extracts the prompts and answers for model tuning.
% \jie{re write this sentence}. 
% Finally, we input the prompts into the generative pre-trained model to generate the input text.
We evaluate the passing rate of text generated by {\tool} on 106 apps of Google Play, and the result shows that the pass rate of {\tool} is 87\%, which is 93\% higher than the best baseline. We further integrate {\tool} with the automated GUI testing tools, and it can cover 42\% more app activities and 52\% more pages compared with the raw tool.
% \jie{should include the integration with gui testing tools, which is important. }

In the future, we will work in two directions.
First, we will improve our approach in extracting the context information in the nearby UI pages to solve the problem of lacking context information of the text input on the current UI page. 
Second, we will not limit {\tool} to app testing, and plan to explore its potential usage in other areas like providing input prompts for the application's auxiliary reading function.

\bibliographystyle{IEEEtran}
	
\bibliography{reference}

\end{document}